\documentclass[a4paper,11pt]{article}
\usepackage{jcappub} 
\usepackage{lineno}
\usepackage{booktabs}
\usepackage{multirow}
\usepackage{longtable}
\usepackage{microtype}
\usepackage{siunitx}
\usepackage[T1]{fontenc}
\usepackage{booktabs}
\usepackage{orcidlink}

\usepackage{xcolor}
\usepackage[normalem]{ulem}

\title{\boldmath Constraints on Dark Energy and Modified Gravity Models from Fast Radio Bursts and Late-Time Geometric Probes}

\author[a,1]{Bruno W. N. Ribeiro\,\orcidlink{0000-0002-6657-263X}}
\author[b,a]{Lázaro L. Sales\,\orcidlink{0000-0002-5352-6642}}
\author[a]{K. E. L. de Farias\,\orcidlink{0000-0002-9418-9566}}
\author[a]{Raiff H. Santos\,\orcidlink{0009-0009-4616-3183}}
\author[a]{Rafael A. Batista\,\orcidlink{0000-0002-7906-1505}}
\author[a]{Amílcar R. Queiroz\,\orcidlink{0000-0002-4785-5589}}
\affiliation[a]{Universidade Federal de Campina Grande,\\ 
R. Aprígio Veloso 882, Bairro Universitário, 58429-900, Campina Grande, PB, Brazil}
\affiliation[b]{Universidade do Estado do Rio Grande do Norte,\\
Av. Prof. Antônio Campos, Pres. Costa e Silva,
59610-210, Mossoró, RN, Brazil}

\emailAdd{brunoweslley92@gmail.com}

\abstract{We investigate the impact of 104 localized FRBs on cosmological
parameter estimation when combined with three established late-time probes:
Cosmic Chronometers (CC), Type Ia Supernovae (SNe), and Baryon Acoustic 
Oscillations (BAO). By performing a Bayesian analysis of three dark energy 
models ($\Lambda$CDM, $w$CDM, and CPL) and three viable $f(R)$ gravity 
scenarios --- the Appleby–Battye (AB), Hu–Sawicki (HS), and Starobinsky (ST)
models ---, we find that FRBs substantially improve the constraints on the
baryon density $\Omega_{\rm b}$ by $25\%$--$43\%$, the Hubble constant $H_0$
by $12\%$--$35\%$, and the SNe absolute magnitude $M_B$ by $10\%$--$32\%$. 
Constraints on dark energy parameters show more modest improvements, with $w$
improving by $\sim 9\%$ in $w$CDM and $(w_0,w_a)$ improving by $\sim(8,22)\%$ in 
the CPL parametrization. Modified gravity parameters remain weakly constrained,
with improvements of only $6\%$--$15\%$, indicating the limited sensitivity of
current datasets to departures from $\Lambda$CDM. The Figure of Merit analysis 
shows overall improvements ranging from $\sim 48\%$ ($\Lambda$CDM) to $\sim 91\%$
(CPL), driven by enhanced precision in the $(H_0, \Omega_{\rm b})$ plane. Model
comparison reveals moderate statistical preference for extensions beyond
$\Lambda$CDM: AIC strongly favors $w$CDM, CPL, HS, and ST with $\Delta\mathrm{AIC} < -7$,
and LRT yields $p \leq 0.004$, while BIC returns to positive evidence
($-3.2 < \Delta\mathrm{BIC} < -2.7$). These results show that FRBs may be useful
as a complementary probe, particularly for constraining $\Omega_{\rm b}$ and
alleviating key late-time degeneracies.}

\begin{document}
\maketitle
\flushbottom

\section{Introduction}
\label{sec:intro}

The discovery of the late-time accelerated expansion of the Universe
\cite{Riess_1998,Perlmutter_1999} has established the cosmological 
constant $\Lambda$ as the simplest explanation for dark energy. However,
the $\Lambda$ Cold Dark Matter ($\Lambda$CDM) paradigm faces profound
theoretical challenges, such as the fine-tuning and cosmic coincidence 
problems \cite{Weinberg_1988}. More recently, this theoretical discomfort
has been amplified by persistent observational tensions that threaten the
internal consistency of the standard model. The most notable is the Hubble
tension, a $4$--$6\sigma$ discrepancy between the $H_0$ value inferred from
the Cosmic Microwave Background (CMB) under $\Lambda$CDM assumptions and
direct local measurements obtained from type Ia Supernovae (SNe~Ia) 
\cite{Di_Valentino_2021,Riess_2022}. This is accompanied by the $S_8$ tension,
in which weak lensing and galaxy clustering surveys find a lower amplitude of
matter fluctuations than predicted by Planck data \cite{Perivolaropoulos_2022}.
These anomalies suggest that the standard model may be an incomplete description
of the dark sector, motivating exploration of extensions beyond $\Lambda$CDM.

Within the General Relativity (GR) framework, a wide class of extensions has been
proposed by relaxing the assumption of a cosmological constant. The simplest
extension is the $w$CDM model, which assumes a constant equation of state (EoS)
parameter $w \neq -1$, while more general parameterizations such as the 
Chevallier--Polarski--Linder (CPL) model allow for time-dependent EoS, describing
dynamical dark energy \cite{Chevallier_2001,Linder_2003}. These models retain the 
geometrical structure of GR while extending the dark sector, and serve as natural
frameworks for testing departures from $\Lambda$CDM while remaining within 
observationally well-tested gravitational theory.

Alternatively, theories of modified gravity based on $f(R)$ functions
offer a geometrically motivated framework that reproduces accelerated 
expansion by modifying the gravitational sector itself, without 
introducing additional exotic components in the energy--momentum tensor. 
In these models, the Ricci scalar $R$ in the Einstein--Hilbert action 
is replaced by a general function of $R$, effectively generating 
acceleration through higher-order curvature corrections rather than 
a dark energy fluid. Several viable $f(R)$ models, including those of 
Appleby \& Battye~\cite{Appleby_2007}, Hu \& Sawicki~\cite{Hu_2007}, 
and Starobinsky~\cite{Starobinsky_2007}, reproduce the observed cosmic 
acceleration while satisfying local gravity constraints and recovering 
GR in the appropriate limit, making them well-motivated targets for 
independent cosmological tests.

Late-time cosmology currently relies on a combination of geometrical
probes, each sensitive to different aspects of the expansion history. 
Cosmic Chronometers (CC) provide direct measurements of the Hubble
parameter through differential age evolution of galaxies, while SNe Ia
act as standardizable candles tracing the luminosity 
distance. Baryon Acoustic Oscillations (BAO), in turn, offer a standard 
ruler that constrains both the angular diameter distance and the Hubble 
rate. Despite their success, these probes exhibit parameter degeneracies
and limited sensitivity to specific components of the cosmic energy budget,
particularly the baryonic sector, highlighting the need for additional 
independent observables.

Recently, in 2007, another astrophysical phenomenon was discovered by 
Lorimer et al. \cite{Lorimer_2007}, the Fast Radio Bursts (FRBs), which are 
millisecond-duration radio transients of  extragalactic origin \cite{Zhang_2023}
whose dispersion measure (DM) encodes information about the integrated 
free-electron content along the line of sight. Since the intergalactic medium
(IGM) contribution to the DM scales with the baryon density and the expansion
history, FRBs with known redshifts may provide a novel and independent cosmological
probe, complementary to CC, SNe, and BAO. Recent observational programs such as 
CHIME/FRB \cite{frb_collaboration_2026} and ASKAP \cite{hotan2021australian} has
dramatically increased the rate of FRB detections, with several hundred events 
now localized to host galaxies~\cite{Chatterjee_2017,Bannister_2019,Prochaska_2019,
Ravi_2019,Bhandari_2020,Bhandari_2023,Ryder_2023,Cassanelli_2024,Li_2026}. 
Early studies have demonstrated the potential of FRBs to constrain the cosmic baryon
density and probe the distribution of ionized gas in the IGM~\cite{Caleb_2023}.
However, current analyses remain limited by systematic uncertainties associated with
host galaxy contributions and line-of-sight inhomogeneities. Since both dark energy
and modified gravity models affect the background expansion history, they leave
imprints on the DM--$z$ relation, making FRBs a complementary probe with particular
sensitivity to the baryonic sector and the expansion history, thus providing an
independent consistency check on existing constraints.

Despite these advances, existing studies on FRBs have focused primarily 
on model-independent cosmographic analyses, $\Lambda$CDM constraints, 
Hubble-constant measurements, or individual dark energy extensions
\cite{Sales_2026,Lemos_2023,Macquart_2020,Caleb_2023,Connor_2025,Liu_2026b}. 
A comprehensive assessment that simultaneously addresses the impact on 
parameter degeneracies across multiple models, the propagation to derived 
cosmological quantities, modified-gravity scenarios, and model selection 
within a unified statistical framework remains lacking. In particular, it is
unclear how FRBs affect the constraining power of joint late-time datasets 
and whether they provide sufficient discriminating power to favor extensions
beyond $\Lambda$CDM when complexity penalties are properly accounted for.

In this work, we perform a systematic and unified investigation of the 
constraining power of FRBs when combined with traditional late-time probes,
which introduces several methodological advances. We simultaneously analyze
both dark energy and viable $f(R)$ modified gravity scenarios within a 
consistent Bayesian framework, quantify improvements on primary
parameters and parameter degeneracies, and apply multiple complementary model
selection criteria to assess statistical  preference. Specifically, our main
contributions are: (i) we perform joint Bayesian analyses for the Baseline
(CC~+~SNe~+~BAO) and Baseline~+~FRB datasets across six cosmological models;
(ii) we quantify the improvement in cosmological parameter constraints through
the reduction of their uncertainties when FRBs are included; (iii) we assess 
the global impact of FRBs using the Figure of Merit (FoM) metric; and (iv) we
evaluate the model selection using the Akaike (AIC) and Bayesian (BIC) information
criteria, as well as the Likelihood Ratio Test (LRT).

This paper is organized as follows. Sec.~\ref{sec:background} introduces 
the cosmological framework and the distance measures relevant for late-time 
probes. The FRB formalism, including the DM modeling and its statistical 
description, is presented in Sec.~\ref{sec:frbs}. In Sec.~\ref{sec:models} 
we describe the cosmological models under consideration, encompassing both 
GR-based dark energy parametrizations and viable $f(R)$ modified gravity 
scenarios. The observational datasets and their respective systematics are 
detailed in Sec.~\ref{sec:data}. In Sec.~\ref{sec:methodology}, we present 
the statistical framework, including the Bayesian inference setup, prior 
choices, joint likelihood construction, numerical implementation, and MCMC 
sampling strategy, together with the Figure of Merit and model selection 
criteria adopted in this work. Our results and discussions are reported in 
Sec.~\ref{sec:results}, and a summary 
of our main conclusions in Sec.~\ref{sec:conclusion}.


\section{Cosmological Background}
\label{sec:background}

This section summarizes the theoretical framework underlying the cosmological 
observables used in this work. We begin by introducing the background dynamics 
of a spatially flat FLRW universe, followed by the definition of the distance 
measures that connect the expansion history to observations.

\subsection{Flat FLRW Universe}
\label{sec:flrw}

We assume a spatially flat, homogeneous and isotropic universe described by the
Friedmann-Lema\^{i}tre-Robertson-Walker (FLRW) metric:
\begin{equation}
    ds^2 = -dt^2 + a^2(t)\left[dr^2 + r^2 d\Omega^2\right],
\label{eq:flrwmetric}
\end{equation}
where $a(t)$ is the scale factor normalized to $a_0 = 1$ today, and 
$H(t) \equiv \dot{a}/a$ is the Hubble parameter with present value $H_0$.
The redshift $z$ is related to the scale factor by $1 + z = 1/a(t)$.

The energy content of the universe is described by a perfect fluid with
energy-momentum tensor:
\begin{equation}
    T_{\mu\nu} = (\rho + p)\,u_\mu u_\nu + p\,g_{\mu\nu},
\label{eq:Tmunu}
\end{equation}
where $\rho$ is the energy density, $p$ is the pressure, and $u^\mu$ is the
four-velocity of the fluid. The component $i$ is characterized by an equation
of state (EoS) $w_i \equiv p_i/\rho_i$, whose energy density evolves as
\begin{equation}
    \rho_i(z) = \rho_{i,0}\,(1+z)^{3(1+w_i)},
\label{eq:rho_evolution}
\end{equation}
for constant $w_i$. The case $w_i=0$ corresponds to pressureless nonrelativistic
matter, such as baryons and cold dark matter, for which $\rho_{\rm m}\propto(1+z)^3$.
On the other hand, $w_i=-1$ corresponds to a cosmological constant, with 
$\rho_\Lambda = \text{const}$. The density parameter is defined as 
$\Omega_i \equiv \rho_{i,0}/\rho_{\rm crit,0}$, where $\rho_{\rm crit,0} = 3H_0^2/8\pi G$
is the critical density today.

\subsection{Distance Measures and Cosmological Observables}
\label{sec:distances}

In a spatially flat FLRW universe, the distance measures relevant to late-time cosmology
are derived from the comoving distance $D_C(z)$, defined as
\begin{equation}
    D_C(z) = c\int_0^z \frac{dz'}{H(z')},
\label{eq:DC}
\end{equation}
which encodes the expansion history through the Hubble rate. The luminosity distance $D_L$
and angular diameter distance $D_A$ follow directly as
\begin{equation}
    D_L(z) = (1+z)\,D_C(z), \qquad 
    D_A(z) = \frac{D_C(z)}{1+z},
\label{eq:DL_DA}
\end{equation}
where $D_L$ accounts for expansion-induced energy loss and time dilation, while $D_A$ relates
the transverse size of a source to its observed angular extent. The $(1+z)$ factor relating $D_L$
and $D_A$ reflects the Etherington reciprocity relation~\cite{Etherington_1933}, which holds in
any metric theory of gravity for photons traveling on null geodesics. 

These two quantities can be obtained by measurement of Supernovae Ia in the case of $D_L(z)$, 
and for $D_A(z)$, it is provided by the baryon acoustic oscillations. The brightness of SNe Ia
is used as a standard candle, and it depends on the luminosity distance to the source. 
The predicted apparent magnitude in the rest-frame $B$-band at redshift $z$ is related to the
distance modulus $\mu_B$ by
\begin{equation}
    m_B(z) = \mu_B(z) + M_B\,, \qquad 
    \mu_B(z) \equiv 5\log_{10}\!\left[
    \frac{D_L(z)}{\mathrm{Mpc}}\right] + 25,
\label{eq:mB}
\end{equation}
where $M_B$ is the absolute magnitude in the  rest-frame $B$-band, treated as a free nuisance
parameter.

The BAO feature is calibrated against the sound horizon at the drag epoch, $r_{\rm d}$, which 
sets the characteristic comoving scale imprinted by acoustic oscillations in the pre-recombination
plasma.  This quantity is defined as the total comoving distance travelled by  sound waves from the
Big Bang until the moment baryons kinematically decouple from the photon fluid at redshift 
$z_{\rm d} \approx 1060$, and it is expressed as:
\begin{equation}
    r_{\rm d} = \int_{z_{\rm d}}^\infty{\frac{c_s(z)}{H(z)}dz}\,,
\end{equation}
where $c_s(z)$ denotes the adiabatic sound speed in the coupled photon-baryon fluid and $z_{\rm d}$
marks the drag epoch, at which the photon pressure becomes dynamically relevant, and the baryonic 
component effectively freezes out of the acoustic oscillations.

Under the  assumption of standard early-universe physics, the resulting fit takes the form~\cite{Adame_2025}
\begin{equation}
    r_{\rm d} = 147.05\left(\frac{\Omega_{\rm b} h^2}{0.02236}\right)^{-0.13}
    \!\!\left(\frac{\Omega_{\rm m} h^2}{0.1432}\right)^{-0.23}
    \!\!\left(\frac{N_{\rm eff}}{3.04}\right)^{-0.1} \mathrm{Mpc},
\label{eq:rd}
\end{equation}
where the pivot values correspond to the Planck best-fit cosmology~\cite{Aghanim_2020}, 
$h \equiv \frac{H_0}{100\,\rm km\,s^{-1}\,Mpc^{-1}}$, and $N_{\rm eff}$ accounts for 
neutrino species that remain ultrarelativistic throughout the pre-recombination epoch. 
In this work, we fix $N_{\rm eff} = 3.046$, corresponding to the standard prediction from
non-instantaneous neutrino decoupling in the early universe.


\section{Fast Radio Bursts}
\label{sec:frbs}

The key observable of an FRB is its Dispersion Measure, which arises from the
frequency-dependent time delay  experienced by the radio signal as it propagates
through ionized plasma~\cite{Petroff_2019}:
\begin{equation}
    \Delta t \propto \left(\frac{1}{\nu_{\rm lo}^2} 
    - \frac{1}{\nu_{\rm hi}^2}\right){\rm DM},
\end{equation}
where $\nu_{\rm lo}$ and $\nu_{\rm hi}$ denote the lower and upper frequencies of
the signal. The DM corresponds to the integrated column density of free electrons
along the line of sight,
\begin{equation}
    {\rm DM} = \int \frac{n_e(z)}{1+z}\,dl\,,
\end{equation}
where the factor $(1+z)^{-1}$ accounts for cosmological time dilation.

The observed DM can be decomposed into distinct physical contributions:
\begin{equation}
    {\rm DM}_{\rm obs}(z) = {\rm DM}_{\rm MW} 
    + {\rm DM}_{\rm IGM} 
    + \frac{{\rm DM}_{\rm host}}{1+z},
\label{eq:DM_tot}
\end{equation}
where ${\rm DM}_{\rm MW} = {\rm DM}_{\rm MW,ISM} + {\rm DM}_{\rm MW,halo}$ 
accounts for the Galactic interstellar medium and halo, ${\rm DM}_{\rm IGM}$
corresponds to the intergalactic medium, and ${\rm DM}_{\rm host}$ represents
the host galaxy contribution. The extragalactic dispersion measure is obtained
by subtracting the Galactic contribution from the observed signal, as
\begin{align}
    {\rm DM}_{\rm ext}(z) 
    &\equiv {\rm DM}_{\rm obs}(z) - {\rm DM}_{\rm MW} \notag \\
    &= {\rm DM}_{\rm IGM} + \frac{{\rm DM}_{\rm host}}{1+z}.
\label{eq:DM_ext}
\end{align}
where the Galactic contribution is estimated using the electron density model
NE2001~\cite{Cordes_2003}. 

The cosmological information encoded in FRBs enters through the contribution of the IGM.
Its ensemble average is given by the Macquart relation~\cite{Macquart_2020},
\begin{equation}
    \langle{\rm DM}_{\rm IGM}(z)\rangle = 
    \frac{3c\,f_{\rm IGM}\,\Omega_b\,H_0^2}{8\pi G m_p}
    \int_0^z 
    \frac{(1+z')\,x_e(z')}{H(z')}\,dz',
\label{eq:DM_IGM}
\end{equation}
where $f_{\rm IGM}$ denotes the fraction of baryons residing in the IGM, $m_p$ is the
proton mass, and $x_e(z)$ is the free-electron fraction per baryon, given by
\begin{equation}
    x_e(z) = Y_H x_{e,H}(z) + \frac{1}{2}Y_{\rm He}
    x_{e,\rm He}(z),
\end{equation}
with hydrogen and helium mass fractions $Y_H = 3/4$ and 
$Y_{\rm He} = 1/4$.

Observational and numerical studies regarding the baryon fraction in the IGM suggest that
the value of $f_{\rm IGM}$ lies within the range $0.7 \lesssim f_{\rm IGM} \lesssim 0.95$ 
\cite{Shull_2012,Wei_2019,Dai_2021,Lemos_2023,Wang_2023,Connor_2025}, reflecting uncertainties
in the baryon distribution between collapsed structures and the diffuse IGM. In this work, we 
adopt $f_{\rm IGM} = 0.83$, a representative value consistent with recent observational and 
simulation-based estimates~\cite{Shull_2012,Dai_2021,Connor_2025}. In addition, since hydrogen 
and helium are fully ionized at $z < 3$ \cite{Meiksin_2009,Becker_2010}, we take $x_{e,H} = x_{e,\rm He} = 1$,
yielding $x_e = 7/8$.

\subsection{Statistical Description of the DM Components}
\label{sec:frb_pdf}

The IGM contribution to ${\rm DM}_{\rm ext}$ is inherently stochastic due to the inhomogeneous
distribution of baryons in the cosmic large-scale structure. The DM accumulated along a given line
of sight traces the integrated column density of free electrons and is therefore sensitive to density
fluctuations along different paths.

Based on semi-analytic models and hydrodynamical simulations, the distribution of ${\rm DM}_{\rm IGM}$
at fixed redshift is approximately Gaussian around its mean, with non-Gaussian tails arising from rare
intersections with dense structures, such as galaxy groups and massive halos, along the line of 
sight~\cite{Macquart_2020,McQuinn_2013,Jaroszynski_2019,Zhang_2021}. More accurate descriptions of the
IGM scatter have been proposed, which include quasi-log-normal profiles~\cite{McQuinn_2013} and 
simulation-calibrated templates~\cite{Jaroszynski_2019}. However, these non-Gaussian corrections become 
significant only at low redshift ($z \lesssim 0.1$), where the small number of intersected halos leads to
a strongly asymmetric DM distribution, and their impact on parameter inference is negligible for the sample
sizes currently available~\cite{Zhang_2021}. We therefore adopt a truncated Gaussian approximation as an
effective description of the IGM scatter, which captures the dominant statistical uncertainty relevant for
the present catalog of 104 FRBs, consistent with the approach adopted in several recent FRB cosmology 
analyses~\cite{Macquart_2020,Zhang_2021}. 

The corresponding probability distribution is written
\begin{equation}
    P_{\rm IGM}({\rm DM}_{\rm IGM}\,|\,z) = 
    \frac{1}{\mathcal{N}(z)}\,
    \frac{1}{\sqrt{2\pi}\,\sigma_{\rm IGM}(z)}
    \exp\!\left[
    -\frac{\left({\rm DM}_{\rm IGM} - \langle {\rm DM}_{\rm IGM}(z)\rangle \right)^2}
    {2\sigma_{\rm IGM}^2(z)}
    \right]
    \Theta({\rm DM}_{\rm IGM}),
\label{eq:P_IGM}
\end{equation}
where $\Theta(x) = 1$ if $x \geq 0$ and $\Theta(x) = 0$ otherwise is the
Heaviside step function enforcing the physical constraint
${\rm DM}_{\rm IGM} \geq 0$, and $\mathcal{N}(z)$ is the normalization
factor of the truncated distribution,
\begin{equation}
    \mathcal{N}(z) = \frac{1}{2}\left[1 + \mathrm{erf}\!\left(
    \frac{\langle {\rm DM}_{\rm IGM}(z)\rangle}{\sqrt{2}\,\sigma_{\rm IGM}(z)}
    \right)\right],
\label{eq:norm_truncated}
\end{equation}
where $\langle {\rm DM}_{\rm IGM}(z)\rangle$ is given by the Macquart relation, 
Eq. \eqref{eq:DM_IGM}, and $\sigma_{\rm IGM}(z)$ encodes the redshift-dependent 
scatter arising from line-of-sight fluctuations in the large-scale structure. 
This truncation eliminates the unphysical probability assigned to negative
dispersion measures by a pure Gaussian. In practice, since
$\langle \mathrm{DM}_{\rm IGM}(z)\rangle \gg \sigma_{\rm IGM}(z)$ for most
sources in our sample, $\mathcal{N}(z) \approx 1$ and the truncated
distribution reduces to the standard Gaussian to high accuracy. Nevertheless,
we adopt the truncated form for statistical consistency, as the correction
becomes non-negligible for low-redshift FRBs where the mean-to-scatter ratio
is smallest.

Following previous studies, we model this scatter empirically 
as a power law in redshift,
\begin{equation}
    \sigma_{\rm IGM}(z) = \sigma_0\,z^{\gamma},
\label{eq:sigma_IGM}
\end{equation}
where $\sigma_0$ and $\gamma$ are coefficients calibrated from simulation-based estimates of 
dispersion measure fluctuations. In particular, we adopt $\sigma_0 = 173.8~{\rm pc\,cm^{-3}}$ 
and $\gamma = 0.4$, following~\cite{Qiang_2020}, where this form is obtained from a power-law 
fit to numerical results describing the impact of large-scale structure on the IGM dispersion.

In contrast to the intergalactic component, the host-galaxy contribution ${\rm DM}_{\rm host}$ 
arises from the local environment of the source and depends on several astrophysical factors,
such as the galaxy morphology, inclination, star formation activity, and the location of the 
burst within the host. These effects introduce significant scatter and intrinsic asymmetry in
the ${\rm DM}_{\rm host}$ distribution. Motivated by cosmological hydrodynamical simulations, 
such as IllustrisTNG, which show that ${\rm DM}_{\rm host}$ follows a log-normal profile across
different redshifts and galaxy types \cite{Zhang_2020}, we model this contribution as
\begin{equation}
    P_{\rm host}({\rm DM}_{\rm host}) = 
    \frac{1}{\sqrt{2\pi}\,\sigma_{\rm host}\,
    {\rm DM}_{\rm host}}
    \exp\!\left[-\frac{1}{2}
    \left(\frac{\ln{\rm DM}_{\rm host} - \mu}{\sigma_{\rm host}}
    \right)^{\!2}\right],
\label{eq:P_host}
\end{equation}
where $e^\mu$ is the median and $\sigma_{\rm host}$ is the logarithmic width, 
both treated as free nuisance parameters.


\section{Cosmological Models}
\label{sec:models}

In this Section, we consider two classes of models that account for the late-time
acceleration: the extensions within General Relativity based on dark energy 
parametrizations and modified theories of gravity that alter the geometric sector
of the field equations.

\subsection{GR-Based Models}

Starting with the framework of GR, it is considered that the late-time accelerated 
expansion of the Universe is given by a dark energy component whose nature remains 
unknown. Hence, we consider $\Lambda$CDM, which is the cosmological standard model 
and two other extensions, $w$CDM and CPL parametrizations.

In the standard cosmological model, it is assumed a spatially flat universe filled 
with pressureless CDM and a cosmological constant $\Lambda$, which acts as a perfect
fluid with EoS $w_\Lambda = -1$ and constant energy density $\rho_\Lambda = \Lambda/(8\pi G)$. 
Using Eq.~(\ref{eq:rho_evolution}), the Friedmann equation takes the form
\begin{equation}
    H^2(z) = H_0^2\left[\Omega_{\rm m}(1+z)^3 
    + \Omega_\Lambda\right],
\label{eq:H_LCDM}
\end{equation}
where $\Omega_{\rm m}$ is the non-relativistic matter density parameter and
$\Omega_\Lambda = 1 - \Omega_{\rm m}$ is the dark energy density parameter. 

The first extension of $\Lambda$CDM replaces the cosmological constant with a 
dark energy component characterized by a constant but free EoS $w \neq -1$. 
From Eq.~(\ref{eq:rho_evolution}), the dark energy density evolves as 
$\rho_{\rm de}(z) \propto (1+z)^{3(1+w)}$, which reduces to a constant for
$w = -1$. The Friedmann equation becomes
\begin{equation}
    H^2(z) = H_0^2\left[\Omega_{\rm m}(1+z)^3 
    + (1-\Omega_{\rm m})(1+z)^{3(1+w)}\right].
\label{eq:H_wCDM}
\end{equation}
The $\Lambda$CDM limit is exactly recovered for $w = -1$, while $w > -1$ ($w < -1$)
corresponds to quintessence (phantom) dark energy.

The Chevallier--Polarski--Linder (CPL) parametrization~\cite{Chevallier_2001,Linder_2003}
provides a minimal and well-motivated extension by allowing $w$ to vary with redshift:
\begin{equation}
    w(z) = w_0 + w_a\frac{z}{1+z} = w_0 + w_a(1-a),
\label{eq:w_CPL}
\end{equation}
where $w_0 \equiv w(z=0)$ is the present-day value and $w_a \equiv -dw/da|_{a=1}$ quantifies 
the rate of change with the scale factor. This parametrization is well-behaved at all redshifts:
it reduces to $w_0$ today and approaches the finite value $w_0 + w_a$ in the early universe 
($z \to \infty$), avoiding the divergences present in some alternative forms. Substituting 
Eq.~(\ref{eq:w_CPL}) into the continuity equation and integrating yields 
$\rho_{\rm de}(z) \propto (1+z)^{3(1+w_0+w_a)}\exp\!\left[-3w_a z/(1+z)\right]$, giving the
Friedmann equation:
\begin{equation}
    H^2(z) = H_0^2\Bigl[\Omega_{\rm m}(1+z)^3 
    + (1-\Omega_{\rm m})(1+z)^{3(1+w_0+w_a)}
    e^{-3w_a z/(1+z)}\Bigr].
\label{eq:H_CPL}
\end{equation}
The $\Lambda$CDM and $w$CDM limits are recovered for $(w_0, w_a) = (-1, 0)$.

\subsection{\texorpdfstring{Modified Gravity: $f(R)$ Theories}{Modified Gravity: f(R) Theories}}
\label{sec:fr_theories}

As a representative class of modified gravity scenarios, 
we focus on $f(R)$ theories, where the Einstein--Hilbert 
action is generalized to a nonlinear function of the 
Ricci scalar. We briefly summarize the general formalism 
and the conditions required for cosmological viability.

\subsubsection{General Framework} 

The modified Einstein--Hilbert action for $f(R)$ gravity
\cite{Sotiriou_2010,De_Felice_2010} reads
\begin{equation}
    S = \int d^4x\sqrt{-g}\left[\frac{M_{\rm Pl}^2}{2}f(R) 
    + \mathcal{L}_{\rm m}\right],
\label{eq:action}
\end{equation}
where $M_{\rm Pl}^2 \equiv (8\pi G)^{-1}$ is the reduced Planck
mass and $\mathcal{L}_{\rm m}$ is the matter Lagrangian density.
Varying Eq.~(\ref{eq:action}) with respect to the metric $g_{\mu\nu}$ 
yields the modified field equations:
\begin{equation}
    f_{,R}R_{\mu\nu} - \frac{f}{2}g_{\mu\nu} 
    - \left(\nabla_\mu\nabla_\nu - g_{\mu\nu}\Box\right)f_{,R} 
    = \frac{T_{\mu\nu}}{M_{\rm Pl}^2},
\label{eq:field_eq}
\end{equation}
where $f_{,R} \equiv df/dR$, and $T_{\mu\nu}$ is the energy-momentum 
tensor of matter. Taking the trace of Eq.~(\ref{eq:field_eq}) gives:
\begin{equation}
    Rf_{,R} - 2f + 3\,\Box f_{,R} = \frac{T}{M_{\rm Pl}^2},
\label{eq:trace}
\end{equation}
where $T \equiv T^\mu{}_\mu$. In a vacuum, Eq.~(\ref{eq:trace}) reduces to
\begin{equation}
    Rf_{,R} - 2f = 0,
\label{eq:vacuum}
\end{equation}
whose positive real roots define the de~Sitter vacuum 
solutions that underpin both the inflationary and the 
late-time accelerated expansion phases of the Universe. 
A necessary condition for the existence of physically
acceptable de~Sitter solutions is
\begin{equation}
    \left.\frac{f_{,R}}{f_{,RR}}\right|_{R=R_*} > R_*,
\label{eq:ds_condition}
\end{equation}
where $f_{,RR} \equiv d^2f/dR^2$ and $R_*$ is a positive real root of 
Eq.~(\ref{eq:vacuum}). This condition ensures the stability of the de~Sitter
solution under perturbations. Moreover, it is closely related to the requirement
that the effective scalar degree of freedom remains well-behaved. It plays a central
role in selecting physically viable $f(R)$ models.

Considering the spatially flat FLRW metric in Eq.~(\ref{eq:flrwmetric}) and the 
perfect-fluid energy-momentum tensor in Eq.~(\ref{eq:Tmunu}), the modified field
equations~(\ref{eq:field_eq}) yield
\begin{equation}
    3f_{,R}H^2 = \rho_{\rm m} + \frac{Rf_{,R} - f}{2} 
    - 3H\dot{f}_{,R},
\label{eq:fr_friedmann1}
\end{equation}
\begin{equation}
    -2f_{,R}\dot{H} = \rho_{\rm m} + \ddot{f}_{,R} 
    - H\dot{f}_{,R},
\label{eq:fr_friedmann2}
\end{equation}
where dots denote derivatives with respect to cosmic time. Radiation is neglected
since we focus on the late-time evolution $z \ll z_{\rm eq} \simeq 3400$, where 
$\rho_{\rm r}(1+z) \ll \rho_{\rm m}$. The Ricci scalar is related to $H$ through 
$R = 6(\dot{H} + 2H^2)$. Together with Eq.~(\ref{eq:field_eq}), these form a closed
system for the background evolution, which reduces to the standard Friedmann 
equations for $f(R) = R - 2\Lambda$.

\subsubsection{Viability conditions}

For a cosmologically viable $f(R)$ model, the following 
conditions must be satisfied simultaneously:
\begin{enumerate}
\item[(i)] \textit{Stability.} The function $f(R)$ must satisfy
\begin{equation}
    f_{,R} > 0, \qquad f_{,RR} > 0
\label{eq:stability}
\end{equation}
over the entire curvature range of cosmological interest. 
The first condition ensures that gravitons are not ghosts 
and that gravity remains attractive; the second prevents
the scalar degree of freedom (\textit{scalaron}) from 
becoming tachyonic in the high-curvature regime.

\item[(ii)] \textit{Newtonian limit.} In the limit $R \gg R_0$, 
where $R_0$ denotes the present-day curvature scalar, 
the model must reduce to GR with small corrections:
\begin{equation}
    |f(R) - R| \ll R, \qquad 
    |f_{,R} - 1| \ll 1, \qquad 
    Rf_{,RR} \ll 1.
\label{eq:newtonian}
\end{equation}
This ensures consistency with the observed small-scale 
matter inhomogeneities and compact object phenomenology 
that are well-described by Newtonian gravity.

\item[(iii)] \textit{Solar System and laboratory constraints.} 
The model must be indistinguishable from GR at the precision
of current laboratory experiments and Solar System tests of 
gravity.

\item[(iv)] \textit{High-curvature GR recovery.} In the 
high-curvature limit, the model must approach GR with a 
cosmological constant:
\begin{equation}
    \lim_{R \gg R_0} f(R) = R - 2\Lambda,
\label{eq:gr_limit}
\end{equation}
which implies $f_{,R}(\infty) = 1$ and therefore 
$0 < f_{,R}(R) < 1$ for all finite $R$. This condition 
also guarantees consistency with CMB observations 
and the standard thermal history of the Universe, 
including Big Bang nucleosynthesis and the 
matter-dominated epoch.

\item[(v)] \textit{Late-time de~Sitter attractor.} The model 
must admit a stable or metastable de~Sitter fixed point 
to account for the observed current phase of accelerated 
expansion.
\end{enumerate}
These conditions ensure that the model is free from 
theoretical pathologies and observationally consistent
across the full range of cosmologically relevant scales. 
In particular, they guarantee the recovery of GR in 
high-curvature regimes, the stability of cosmological 
solutions, and a viable late-time accelerated phase.
Together, they define the physically admissible 
parameter space of $f(R)$ models.

\subsubsection{Starobinsky regularization}

A generic issue in viable $f(R)$ models is the emergence of a weak curvature
singularity when $f_{,RR} \to 0$ at finite $R$, marginally violating the 
stability condition $f_{,RR} > 0$. In this regime, the scalaron mass 
$m_\phi^2 \sim 1/(3f_{,RR})$ diverges, leading to pathological behavior such
as large-amplitude curvature oscillations in high-curvature environments 
($R \gg R_{\rm vac}$). This problem can be solved by adding a quadratic correction
proportional to $R^2$, which guarantees $f_{,RR} > 0$ at all curvatures while 
remaining subdominant at late times. We therefore implement the regularization
\cite{Starobinsky_2007,De_Felice_2010}
\begin{equation}
    f(R) \rightarrow f(R) + \frac{R^2}{6M^2}\,, 
    \qquad M^2 \equiv \frac{\epsilon_{f(R)}}{\delta_s}\,,
\end{equation}
where $\epsilon_{f(R)}$ is the characteristic curvature scale of the model at the present
epoch, $\delta_s \ll 1$ is a dimensionless parameter and $M$ characterizes a mass scale 
coinciding with the scalaron rest-mass whenever low curvature modifications to GR can be 
neglected. To reproduce the observed amplitude of the primordial power spectrum, one typically
requires $M \approx 1.5 \times 10^{-5} (50/N) M_{\rm Pl}$ at the end of inflation~\cite{Appleby_2010}.

\subsubsection{Appleby-Battye}
\label{sec:ab}

The Appleby--Battye (AB) model, originally proposed in~\cite{Appleby_2007}, is defined as
\begin{equation}
    f(R) = \frac{R}{2} - \frac{\epsilon_{\rm AB}}{2}\ln\!\left[\frac{\cosh\!\left(
    \frac{R}{\epsilon_{\rm AB}} - b\right)}{\cosh(b)}\right],
\label{ab:model}
\end{equation}
where $b$ is a dimensionless parameter controlling the deviation from $\Lambda$CDM, and 
$\epsilon_{\rm AB}$ sets the characteristic curvature scale. The latter is fixed by the 
de~Sitter vacuum condition $f(R_{\rm vac}) = R_{\rm vac}$:
\begin{equation}
    \epsilon_{\rm AB} = \frac{R_{\rm vac}}{b + \ln\!\left(2\cosh b\right)},
\end{equation}
with $R_{\rm vac} = 12H_0^2$. The stability condition $b > 1.6$ must be satisfied in order
for the model to admit a viable late-time de~Sitter solution and reproduce the observed 
accelerated expansion of the Universe \cite{Appleby_2010,Motohashi_2012,Nishizawa_2014,Ribeiro_2023}. 
In the limit $b \to \infty$, as well as for high curvatures $R \gg R_{\rm vac}$, the model
approaches $\Lambda$CDM.

\subsubsection{Hu-Sawicki}
\label{sec:hs}

Originally proposed in~\cite{Hu_2007}, the Hu-Sawicki (HS) model can be written 
as\footnote{The original form of the HS model is 
$f(R) = R - m^2 \frac{c_1(R/m^2)^n}{c_2(R/m^2)^n + 1}$, where
$m^2 \equiv \frac{\kappa^2\rho_0}{3} = H_0^2\Omega_{\rm m}$. The parametrization
in Eq.~(\ref{hs:model}) is obtained by defining $\mu_{\rm HS}^2 \equiv m^2\,c_2^{-1/n}$
and identifying $\Lambda = \frac{m^2 c_1}{2c_2}$.}:
\begin{equation}
    f(R) = R - 2\Lambda\frac{R^n}{R^n + \mu_{\rm HS}^{2n}},
\label{hs:model}
\end{equation}
where $\Lambda = 3H_0^2(1-\Omega_{\rm m})$ is the effective cosmological constant and $n$
is a positive integer. For $R \gg \mu_{\rm HS}^2$, the model approaches $\Lambda$CDM, and
it is exactly recovered in the limit $\mu_{\rm HS} = 0$. In contrast, for 
$\mu_{\rm HS} \to \infty$ the model reduces to $f(R) = R$, corresponding to GR without a 
cosmological constant and therefore to a non-accelerating universe.

\subsubsection{Starobinsky}
\label{sec:st}

Proposed in~\cite{Starobinsky_2007} as a geometric dark energy model, the Starobinsky (ST)
model is defined by:
\begin{equation}
    f_{\rm S}(R) = R + \lambda_{\rm S} R_{\rm S} 
    \left[\left(1 + \frac{R^2}{R_{\rm S}^2}\right)^{-n} - 1\right] \,,
\label{st:model}
\end{equation}
where $\lambda_{\rm S} > 0$ and $R_{\rm S} \sim R_0$ sets the characteristic curvature scale.
As in the HS model, $n$ is a positive integer. In the high-curvature regime $R \gg R_{\rm S}$,
the model approaches $\Lambda$CDM with an effective cosmological constant 
$\Lambda = \lambda_{\rm S} R_{\rm S}/2$. Imposing consistency with the background expansion
today yields
\begin{equation}
    R_{\rm S} = \frac{6 H_0^2 (1 - \Omega_{\rm m})}{\lambda_{\rm S}}\,.
\label{starob.vinc}
\end{equation}
Furthermore, the requirement of a stable late-time de~Sitter solution imposes a lower bound
on $\lambda_{\rm S}$ that depends on $n$. For representative values, one finds 
$(n, \lambda_{\rm S,min}) \simeq (1,\,1.54)$, $(2,\,0.94)$, $(3,\,0.73)$, and 
$(4,\,0.61)$~\cite{Motohashi_2010}. Consequently, only a restricted region of the parameter
space $(\lambda_{\rm S}, n)$ corresponds to physically viable cosmological solutions.


\section{Observational Datasets}
\label{sec:data}

We constrain the free parameters of all models described in Sec.~\ref{sec:models} using four
independent late-time datasets composed of 35 $H(z)$ measurements from CC, 1590 light curves
from the Pantheon$+$ SNe~Ia compilation (after a $z > 0.01$ redshift cut), 13 BAO measurements
from DESI~DR2, and 104 FRBs with measured redshifts. All distance observables entering the 
likelihoods are defined in Sec.~\ref{sec:distances}.

\subsection*{Cosmic Chronometers}
\label{sec:cc}

A well-established approach for probing the expansion history of the Universe in a 
model-independent way is the Cosmic Chronometers method. This technique relies on the
relation $H(z) = -\frac{1}{1+z}\frac{dz}{dt} \approx -\frac{1}{1+z}\frac{\Delta z}{\Delta t}$.
The measurement of the CC dataset is inferred from pairs of passively evolving galaxies
with old stellar populations, negligible star formation, and slightly different redshifts~\cite{Jimenez_2002}.

In Table~\ref{tab:cc} we list 35 measurements of $H(z)$, which carry systematic 
uncertainties mainly associated with stellar population synthesis model (SPS) 
modeling and potential contamination from residual young stellar populations in
the galaxies~\cite{Valent_2019,Yang_2020}.

\subsection*{Type Ia Supernovae}
\label{sec:sne}

We use the Pantheon$+$ compilation~\cite{Scolnic_2022}, comprising 1701 light curves
of 1550 spectroscopically confirmed SNe~Ia over $0.001 \leq z \leq 2.26$, with a full
statistical and systematic covariance matrix $C_{\rm SNe}$. We apply a low-redshift 
cut $z > 0.01$ to minimize contamination from peculiar velocities, which at very low redshifts 
become comparable to the Hubble flow and introduce systematic biases in the inferred 
luminosity distances~\cite{Davis_2011}. 
The nuisance parameter $M_B$ is treated as a free parameter and constrained jointly
with the cosmological parameters, with the uniform prior listed in Table~\ref{tab:priors}.

\subsection*{Baryon Acoustic Oscillations}
\label{sec:bao}

BAO provides one of the most robust standard rulers in cosmology, enabling precise measurements
of the expansion history of the Universe. This characteristic comoving scale originates from 
sound waves propagating in the photon--baryon plasma before recombination, and is set by the 
sound horizon at the drag epoch, $r_d$.

Here, we use 13 BAO measurements from DESI DR2~\cite{Abdul_Karim_2025}, which includes over 14 
million galaxies and quasars spanning the redshift range $0.1 < z < 4.2$. The dataset comprises 
multiple tracers, including Bright Galaxy Samples (BGS), Luminous Red Galaxies (LRGs), Emission 
Line Galaxies (ELGs), quasars (QSOs), and Ly$\alpha$ forest measurements, probing a wide range of
cosmic epochs. All observables are expressed in units of the sound horizon at the drag epoch, $r_d$,
including measurements of $D_M(z)/r_d$, $D_H(z)/r_d$, and $D_V(z)/r_d$, with $r_d$ calculated according
to Eq.~(\ref{eq:rd}).

\subsection*{Fast Radio Bursts}
\label{sec:data_frb}

We use 104 FRBs with spectroscopically confirmed host-galaxy redshifts, compiled from 
a heterogeneous set of recent surveys and follow-up observations. The sample spans the
redshift range $0.0085 \leq z \leq 1.354$, and originates from multiple radio telescope
facilities, including CHIME~\cite{Amiri_2025}, ASKAP/CRAFT~\cite{Shannon_2025}, and 
DSA~\cite{Sharma_2024,Law_2024,Connor_2025}, among others 
\cite{Bhardwaj_2024,Gordon_2023,Heintz_2020,Bhandari_2022,Michilli_2023,Ibik_2024,Ravi_2022,Chatterjee_2017,Ryder_2023},
as shown in Table~\ref{tab:frbs}.

The Galactic interstellar contribution ${\rm DM}_{\rm MW,ISM}$ is estimated using the NE2001
electron density model \cite{Cordes_2003}, spanning $19.9$--$188.4~{\rm pc\,cm^{-3}}$ across
the sample. The halo contribution ${\rm DM}_{\rm MW,halo}$ remains uncertain, with estimates 
ranging from $\sim 25$--$110~{\rm pc\,cm^{-3}}$ depending on the assumed gas distribution and
feedback mechanisms \cite{Yamasaki_2020, Cook_2023,Keating_2020,Platts_2020}, and potentially
direction-dependent in anisotropic models \cite{Liu_2026}.
Following~\cite{Macquart_2020,Prochaska_2019,Sales_2026}, we adopt the fiducial constant value
${\rm DM}_{\rm MW,halo} = 50~{\rm pc\,cm^{-3}}$, which is representative of current constraints
and consistent with the expected baryon fraction of the Milky Way. It is worth stressing that we
use both repeating and non-repeating sources in our MCMC routine.


\section{Statistical Methodology}
\label{sec:methodology}

We perform a Bayesian inference analysis to constrain the cosmological parameters
of each model. According to Bayes' theorem, the posterior probability distribution
is given by
\begin{equation}
    P(\boldsymbol{\theta}|D) \propto 
    \mathcal{L}(D|\boldsymbol{\theta})\,\pi(\boldsymbol{\theta}),
\end{equation}
where $\boldsymbol{\theta}$ denotes the set of model parameters, $D$ is the observed
data, $\mathcal{L}$ is the likelihood function, and $\pi(\boldsymbol{\theta})$ is the
prior distribution. 

We adopt a combination of uniform (top-hat) priors implemented as hard bounds, together
with Gaussian priors for selected parameters, reflecting either minimal assumptions or
external constraints from independent observations. Table~\ref{tab:priors} summarizes 
the prior distributions adopted for all free parameters.

\begin{table}[htbp]
\centering
\begin{tabular}{lcc}
\toprule
\textbf{Parameter} & \textbf{Prior type} & \textbf{Range} \\
\midrule
$H_0$ [km\,s$^{-1}$\,Mpc$^{-1}$] & Uniform & $[50,\,90]$ \\
$\Omega_{\rm m}$                 & Uniform & $[0.1,\,0.6]$ \\
$\Omega_{\rm b}$                 & Uniform & $[0.01,\,0.1]$ \\
$M_B$ [mag]                      & Uniform & $[-20.0,\,-18.0]$ \\
\midrule
$w$     & Uniform & $[-3,\,1]$ \\
$w_0$   & Uniform & $[-3,\,1]$ \\
$w_a$   & Uniform & $[-3,\,2]$ \\
\midrule
$b$                 & Uniform & $[1.6,\,12]$ \\
$\mu_{\rm HS}$ [km\,s$^{-1}$\,Mpc$^{-1}$] & Uniform & $[0,\,300]$ \\
$\lambda_{\rm S}$   & Uniform & $[1.54,\,10]$ \\
\midrule
$\sigma_{\rm host}$ & Uniform & $[0.2,\,2.0]$ \\
$e^\mu$ [pc\,cm$^{-3}$] & Uniform & $[20,\,200]$ \\
\bottomrule
\end{tabular}
\caption{Prior distributions adopted for all free parameters.
Uniform priors are specified by their bounds $[a,b]$.}
\label{tab:priors}
\end{table}

For the primary cosmological parameters ($H_0$, $\Omega_{\rm m}$,
$\Omega_{\rm b}$), we assign broad uniform priors so that the constraints
are driven entirely by the data; in particular, the range adopted for
$\Omega_{\rm b}$ is physically motivated and broad enough to allow it to
be constrained directly by the BAO and FRB data, which are sensitive to
the baryon density through the sound horizon and the IGM dispersion
measure, respectively. For the SNe~Ia absolute magnitude $M_B$, we assign
a broad uniform prior, allowing it to be constrained jointly with all the
cosmological parameters. Finally, for the extended-model parameters ($w$,
$w_0$, $w_a$, $b$, $\mu_{\rm HS}$, $\lambda_{\rm S}$) and the FRB
nuisance parameters ($e^\mu$, $\sigma_{\rm host}$), we adopt uniform
priors that are minimally informative and comfortably cover the regions
of interest suggested by previous studies and exploratory runs.

\subsection{Numerical Implementation}
\label{sec:numerics}

For the GR-based cosmological models ($\Lambda$CDM, $w$CDM, CPL), the Hubble parameter
$H(z)$ is computed analytically at each likelihood evaluation. In contrast, the background
evolution in $f(R)$ gravity is governed by higher-order differential systems compared to GR.
Depending on the structure of each model, the modified Friedmann equations can be recast 
into equivalent dynamical systems for the background expansion, which are then integrated 
numerically (see Appendix~\ref{app:ode} for details).

The numerical integration strategy is adapted to the specific structure of each model.
For the AB case, whose background evolution is expressed as a third-order ordinary 
differential equation (ODE) for $H(a)$ and remains numerically well behaved over the 
relevant redshift range, we use the explicit high-order Runge--Kutta solver DOP853, 
which is well suited for smooth, non-stiff systems.  In contrast, for the HS and ST 
models, which are formulated as second-order systems in the variable $y_H(\ln a)$ and
can exhibit stiffness due to the scalaron dynamics at low curvature, we employ the 
LSODA solver~\cite{Hindmarsh_1983}, which automatically switches between stiff and 
non-stiff integration schemes.

Since repeated ODE integrations during the MCMC sampling would be
computationally prohibitive, we adopt a two-step strategy. First, the
background equations are solved over a precomputed grid in
$(H_0,\,\Omega_{\rm m},\,\theta_{f(R)},\,z)$, covering the redshift range
$z \in [0,\,4]$. Second, during sampling, $H(z)$ is evaluated via
interpolation over this grid, reducing each likelihood call to an
inexpensive lookup. Full details of the grid construction and validation
are given in Appendix~\ref{app:grid}.

\subsection{Joint Likelihood} 
\label{sec:likelihood}

Assuming statistical independence between the different datasets (while fully 
accounting for internal covariances within each dataset), the total log-likelihood
is constructed as the sum of the individual contributions:
\begin{equation} 
    \ln\mathcal{L}_{\rm tot} = \ln\mathcal{L}_{\rm CC} 
    + \ln\mathcal{L}_{\rm SNe} + \ln\mathcal{L}_{\rm BAO} 
    + \ln\mathcal{L}_{\rm FRB}.
\end{equation}

The CC likelihood is adopted as a Gaussian:
\begin{equation}
    \ln\mathcal{L}_{\rm CC} = -\frac{1}{2}\sum_i\left[ \frac{H_{\rm obs}(z_i) 
    - H_{\rm th}(z_i)}{\sigma_{H,i}} \right]^2,
\end{equation}
where $H_{\rm obs}(z_i)$ and $\sigma_{H,i}$ are the observed Hubble parameter
and its uncertainty at redshift $z_i$, and $H_{\rm th}(z_i)$ is the model prediction.

The SNe likelihood accounts for the full inter-supernova covariance:
\begin{equation}
    \ln\mathcal{L}_{\rm SNe} = -\frac{1}{2}\,\boldsymbol{\Delta m}_B^T\, C_{\rm SNe}^{-1}\,\boldsymbol{\Delta m}_B,
\end{equation}
where $\boldsymbol{\Delta m}_B = \mathbf{m}_{B,\rm obs} 
- \mathbf{m}_{B,\rm th}$ and $\mathbf{m}_{B,\rm th}$ is 
evaluated via Eq.~(\ref{eq:mB}). 

The BAO likelihood is analogously constructed using the full covariance matrix:
\begin{equation}
    \ln\mathcal{L}_{\rm BAO} = -\frac{1}{2}\,\boldsymbol{\Delta d}^T\, C_{\rm BAO}^{-1}\,\boldsymbol{\Delta d},
\end{equation}
where $\boldsymbol{\Delta d} = \mathbf{d}_{\rm obs} - 
\mathbf{d}_{\rm th}$ and $\mathbf{d}_{\rm th}$ is the model prediction for
the BAO observables. 

The FRB likelihood is constructed by marginalizing over the 
unobserved host-galaxy contribution. Given the decomposition of 
Eq.~(\ref{eq:DM_ext}), the probability of measuring 
${\rm DM}_{{\rm ext},i}$ at known redshift $z_i$ is obtained 
by convolving $P_{\rm IGM}$ and $P_{\rm host}$:
\begin{equation}
    \ln\mathcal{L}_{\rm FRB} = \sum_i \ln P\!\left( {\rm DM}_{{\rm ext},i}\,\big|\,z_i\right),
\end{equation}
with the marginal probability density function (PDF)
\begin{equation} 
    P({\rm DM}_{\rm ext}|z) = \frac{1}{\langle{\rm DM}_{\rm IGM}\rangle} \int_0^{{\rm DM}_{\rm ext}(1+z)}\!\!\! P_{\rm IGM}\!\left( \frac{{\rm DM}_{\rm ext} - (1+z)^{-1}{\rm DM}_{\rm host}} {\langle{\rm DM}_{\rm IGM}\rangle} \right)\! P_{\rm host}({\rm DM}_{\rm host}) \, d\,{\rm DM}_{\rm host}, 
\label{eq:P_DMext}
\end{equation}
where $P_{\rm IGM}$ and $P_{\rm host}$ are defined in 
Eqs.~(\ref{eq:P_IGM}) and~(\ref{eq:P_host}), and the 
normalization factor $\langle{\rm DM}_{\rm IGM}\rangle^{-1}$
ensures that the PDF of ${\rm DM}_{\rm ext}$ integrates to
unity~\cite{Macquart_2020, Zhang_2020, McQuinn_2013}.

\subsection{Figure of Merit}
\label{sec:fom}

To quantify the impact of FRBs on the cosmological constraints, 
we evaluate the Figure of Merit (FoM) in the two-dimensional 
parameter space $(H_0,\Omega_{\rm b})$. This choice is physically
motivated by the direct sensitivity of FRB dispersion measures to
the cosmic baryon content, $\Omega_{\rm b}$, which provides 
information complementary to the standard sound-horizon constraints
from BAO observations.

Since some of the posterior distributions obtained in this work exhibit
mild deviations from Gaussianity, particularly in extended cosmological
models, we adopt a non-parametric definition of the FoM based on the 
highest-posterior-density (HPD) region of the joint posterior distribution.
The two-dimensional posterior is reconstructed using a Gaussian kernel 
density estimator (KDE), and the FoM is defined as
\begin{equation}
\mathrm{FoM}_{\rm HPD} = \frac{1}{A_{68.3}},
\end{equation}
where $A_{68.3}$ denotes the area enclosed by the 68.3\% HPD contour in the
$(H_0,\Omega_{\rm b})$ plane.

The HPD contour is determined by identifying the density threshold that encloses
68.3\% of the total posterior probability. Consequently, the FoM directly measures
the constraining power of a given dataset through the inverse area of the corresponding
credible region. Smaller credible regions therefore correspond to larger FoM values.

This definition does not assume Gaussian or elliptical posterior distributions and 
naturally reduces to the standard covariance-based FoM for approximately Gaussian 
posteriors. As a result, the adopted estimator provides a robust measure of the 
constraining power even in the presence of asymmetric or mildly non-Gaussian posterior
distributions.

For completeness, we additionally report effective marginal uncertainties derived from
the 16th and 84th percentiles of the posterior samples, together with the skewness of 
the one-dimensional distributions. These quantities serve as diagnostics of the posterior
shape but do not enter the definition of the FoM itself.

\subsection{Model Selection Criteria}
\label{sec:selection}

To compare models with different numbers of free parameters,
we employ information criteria that penalize model complexity,
following the principle of parsimony. The AIC and BIC are 
defined as
\begin{equation}
    {\rm AIC} = 2k - 2\ln\mathcal{L}_{\max}, \qquad
    {\rm BIC} = k\ln N_{\rm tot} - 2\ln\mathcal{L}_{\max},
\end{equation}
where $k$ is the number of free parameters, 
$N_{\rm tot}$ is the total number of data points, 
and $\mathcal{L}_{\max}$ is the maximum likelihood. 
We report differences $\Delta{\rm AIC}$ and 
$\Delta{\rm BIC}$ relative to $\Lambda$CDM, which 
serves as the baseline model. The interpretation of 
$\Delta{\rm IC} \equiv \Delta{\rm AIC}$ or
$\Delta{\rm BIC}$ follows~\cite{plaza_2025}, where a negative
(positive) $\Delta{\rm IC}$ indicates preference for the alternative
(reference) model, with strength classified as:
\begin{itemize}
    \item $|\Delta{\rm IC}| \in [0,\,2]$: weak evidence, inconclusive;
    \item $|\Delta{\rm IC}| \in (2,\,6]$: positive evidence;
    \item $|\Delta{\rm IC}| \in (6,\,10]$: strong evidence;
    \item $|\Delta{\rm IC}| > 10$: very strong evidence.
\end{itemize}

All models considered in this work are extensions of $\Lambda$CDM. Therefore,
$\Lambda$CDM can be recovered as a limiting case for specific values of the
additional parameters. For this reason, we additionally apply the likelihood-ratio
test (LRT) as a complementary model-comparison criterion. Under the null hypothesis
that $\Lambda$CDM is the true model, the statistic
\begin{equation}
    \Lambda_{\rm LR} = -2\ln\frac{\mathcal{L}_{\Lambda{\rm CDM}}}
    {\mathcal{L}_{\rm alt}}
\end{equation}
follows asymptotically a $\chi^2$ distribution with $\Delta k$ degrees of 
freedom~\cite{Wilks_1938}. The $p$-value is then obtained directly from this
distribution as
\begin{equation}
    p = P\!\left(\chi^2_{\Delta k} \geq \Lambda_{\rm LR}^{\rm obs}
    \right) = 1 - F_{\chi^2_{\Delta k}}\!\left(
    \Lambda_{\rm LR}^{\rm obs}\right),
\end{equation}
where $F_{\chi^2_{\Delta k}}$ is the cumulative distribution function of the
$\chi^2$ distribution with $\Delta k$ degrees of freedom. A value of $p < 0.05$
indicates that the improvement in fit is statistically significant and that the
extra parameters are warranted by the data. On the other hand, $p \geq 0.05$ 
suggests that the additional complexity is not justified relative to $\Lambda$CDM 
model.


\section{Results and discussions}
\label{sec:results}

The results of the observational constraints for all six cosmological scenarios
considered in Sec.~\ref{sec:models} are now presented, where we quantify the 
impact of incorporating FRB observations into standard cosmological probes and 
evaluate the relative performance of each model using model selection criteria 
already discussed. For clarity, we define two data combinations used throughout
this work: the Baseline dataset, corresponding to CC~+~SNe~+~BAO, and the 
Baseline~+~FRB dataset, corresponding to CC~+~SNe~+~BAO~+~FRB. Our analysis
focuses on precision gains, degeneracy breaking, the constraining power of FRBs, 
and the statistical comparison among the different cosmological scenarios analyzed.

\subsection{\texorpdfstring{$\Lambda$CDM}{LCDM}}
\label{sec:results_lcdm}

\begin{figure}
\centering
\includegraphics[width=0.7\linewidth]{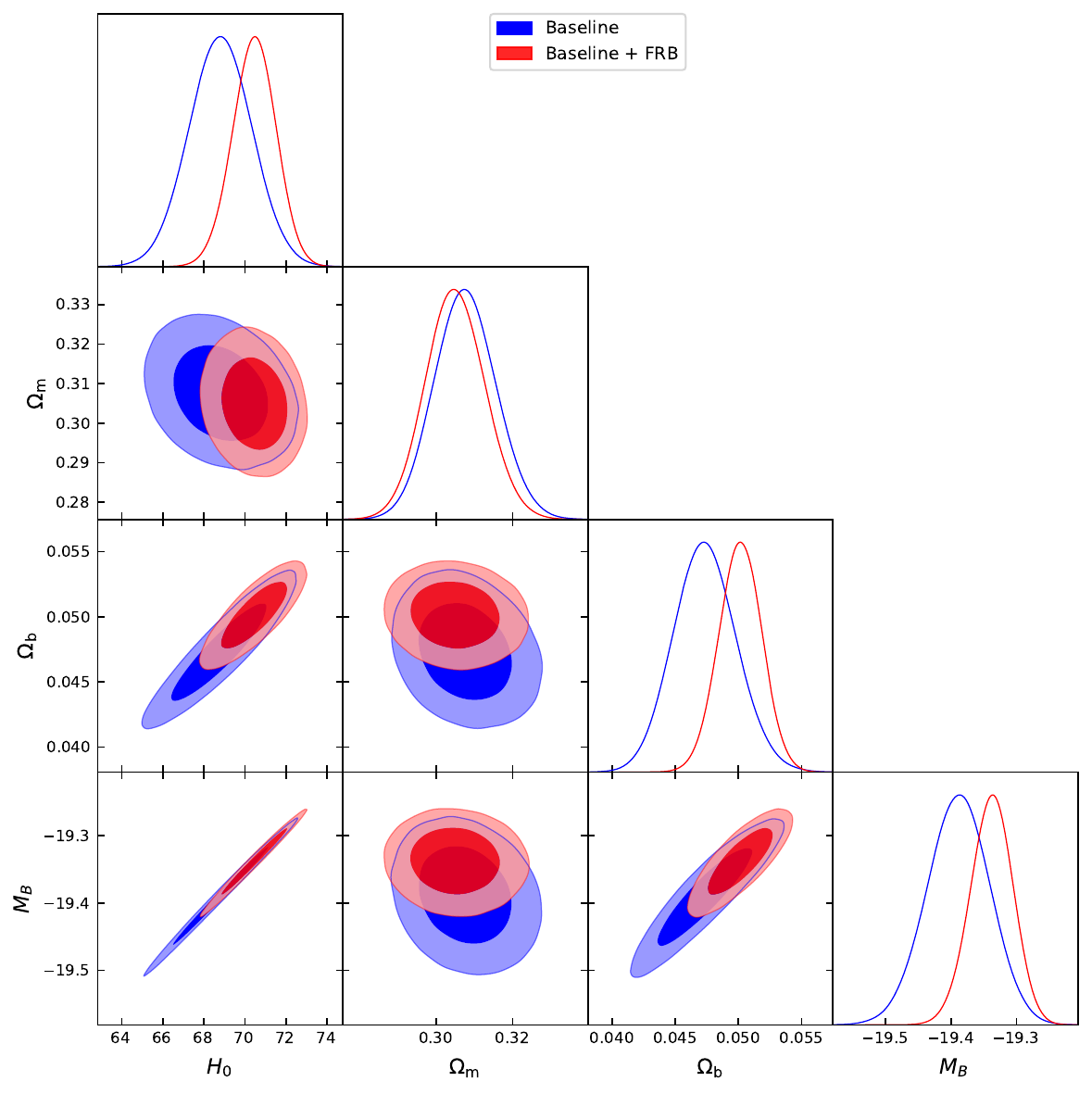}
\caption{Posterior distributions for the $\Lambda$CDM model for the Baseline dataset
(blue) and including FRBs (red). Contours correspond to the 68\% and 95\% credible regions.}
\label{fig:lcdm_corner}
\end{figure}

The posterior distributions for the $\Lambda$CDM model are shown in 
Fig.~\ref{fig:lcdm_corner}, while the corresponding marginalized constraints
are summarized in Table~\ref{tab:lcdm_constraints}. The Baseline dataset 
already provides tight constraints on the expansion history. The inclusion
of FRB observations, however, significantly enhances the overall precision.
In particular, the uncertainty in the Hubble constant $H_0$ decreases from 
$\sim 1.53$ to $\sim 1.04$ km,s$^{-1}$,Mpc$^{-1}$ (a $31.1\%$ improvement),
the uncertainty in the baryon density $\Omega_{\rm b}$ decreases from 
$\sim 0.0025$ to $\sim 0.0017$ (a $30.6\%$ improvement), and the uncertainty
in the SNe~Ia absolute magnitude $M_B$ decreases from $\sim 0.047$ to 
$\sim 0.032$ (a $31.6\%$ improvement). In contrast, the matter density parameter
$\Omega_{\rm m}$ remains essentially unchanged, as it is already tightly constrained
by geometric probes and only weakly correlated with the FRB observable.

These improvements are clearly illustrated in Fig.~\ref{fig:lcdm_corner}, where 
the inclusion of FRB data significantly alleviates the $H_0$--$M_B$ and 
$H_0$--$\Omega_{\rm b}$ degeneracies. This effect is evidenced by the contraction
and partial rotation of the posterior contours, indicating that FRBs provide 
additional information that is largely complementary to the Baseline dataset. 
Consequently, the allowed parameter volume is substantially reduced, making the
$\Lambda$CDM model a useful reference case for assessing the impact of FRBs on 
more extended cosmological scenarios.

\begin{table}
\centering
\begin{tabular}{lcc}
\toprule
\textbf{Parameter}  & \textbf{Baseline}            & \textbf{Baseline + FRB} \\
\midrule
$H_0$               & $68.81^{+1.53}_{-1.52}$      & $70.46^{+1.04}_{-1.06}$       \\
$\Omega_{\rm m}$    & $0.308 \pm 0.008$            & $0.305 \pm 0.008$             \\
$\Omega_{\rm b}$    & $0.0473^{+0.0025}_{-0.0024}$ & $0.0501 \pm 0.0017$           \\
$M_B$               & $-19.388^{+0.047}_{-0.048}$  & $-19.338^{+0.032}_{-0.033}$   \\
\midrule
$\sigma_{\rm host}$ & ---                          & $0.833^{+0.124}_{-0.102}$     \\
$e^\mu$             & ---                          & $104.2^{+16.2}_{-15.8}$       \\
\bottomrule
\end{tabular}
\caption{Marginalized constraints at 68\% confidence level for the $\Lambda$CDM model,
obtained from the baseline dataset and from the combined dataset including FRBs.}
\label{tab:lcdm_constraints}
\end{table}

\subsection{\texorpdfstring{$w$CDM}{wCDM}}
\label{sec:results_wcdm}

\begin{figure}
\centering
\includegraphics[width=0.8\linewidth]{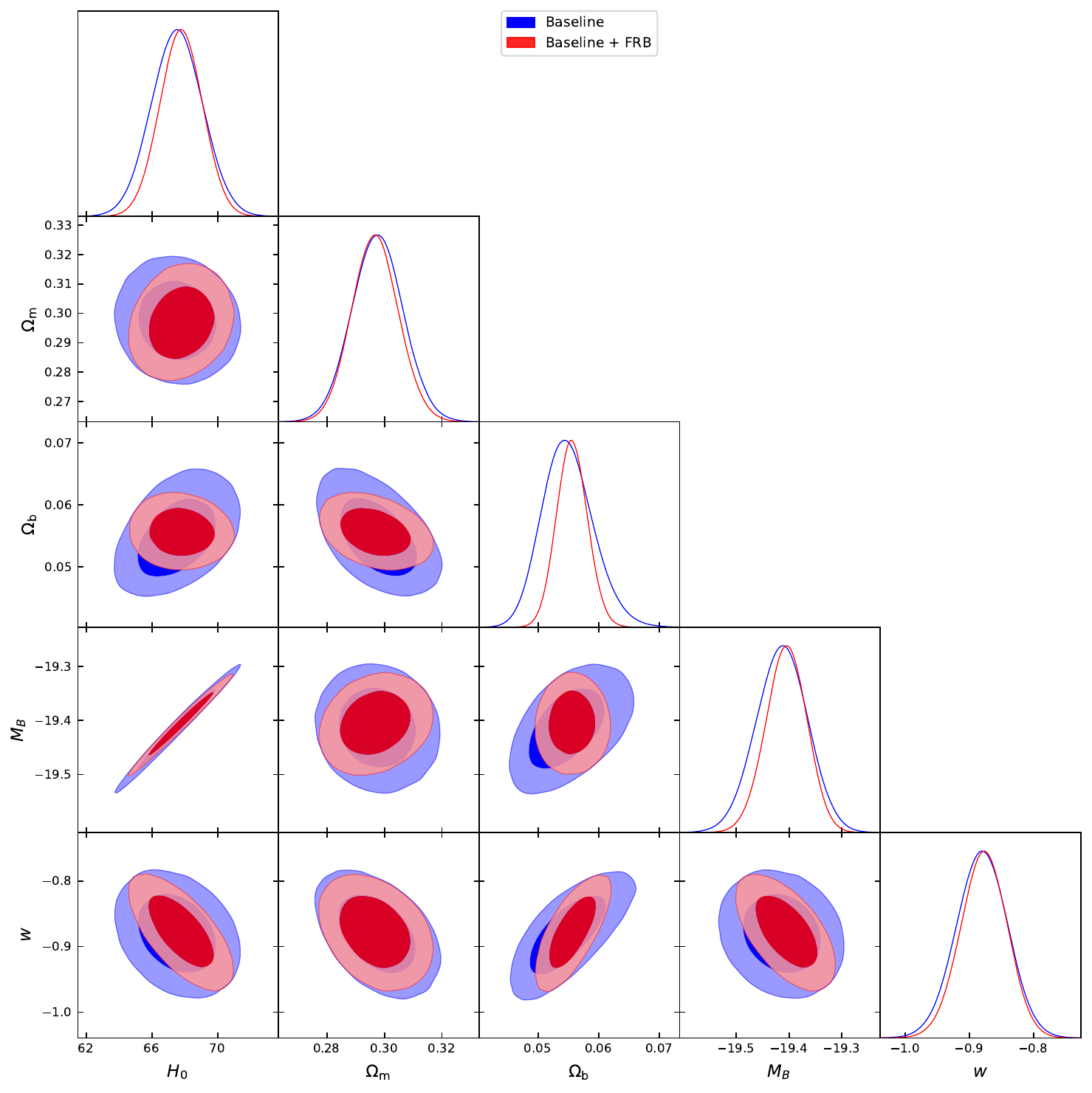}
\caption{Posterior distributions for the $w$CDM model for the Baseline dataset (blue) and including
FRBs (red). Contours correspond to the 68\% and 95\% credible regions.}
\label{fig:wcdm_corner}
\end{figure}

The posterior distributions for the $w$CDM model are shown in Fig.~\ref{fig:wcdm_corner},
with the corresponding marginalized constraints listed in Table~\ref{tab:wcdm_constraints}. 
Allowing for a free dark energy EoS parameter introduces additional freedom in the expansion
history, leading to stronger degeneracies between the expansion rate and the dark energy sector.
The inclusion of FRBs nevertheless improves the constraints on several cosmological parameters.
In particular, the uncertainty in the Hubble constant $H_0$ decreases from $\sim 1.56$ to 
$\sim 1.30$ km,s$^{-1}$,Mpc$^{-1}$ (a $16.4\%$ improvement), while the uncertainty in the baryon 
density parameter $\Omega_{\rm b}$ decreases from $\sim 0.0041$ to $\sim 0.0026$ (a $37.8\%$ 
improvement). The uncertainty in the SNe~Ia absolute magnitude $M_B$ is also reduced from 
$\sim 0.049$ to $\sim 0.039$ (a $20.6\%$ improvement). In contrast, the dark energy parameter $w$
exhibits only a modest reduction in its uncertainty, from $\sim 0.040$ to $\sim 0.036$ (an $8.9\%$ 
improvement).

\begin{table}
\centering
\begin{tabular}{lcc}
\toprule
\textbf{Parameter}  & \textbf{Baseline}      & \textbf{Baseline + FRB} \\
\midrule
$H_0$                & $67.51^{+1.56}_{-1.55}$      & $67.75 \pm 1.30$       \\
$\Omega_{\rm m}$     & $0.298 \pm 0.009$            & $0.297 \pm 0.008$             \\
$\Omega_{\rm b}$     & $0.0548^{+0.0043}_{-0.0039}$ & $0.0556^{+0.0026}_{-0.0025}$    \\
$M_B$                & $-19.413^{+0.048}_{-0.049}$  & $-19.405^{+0.038}_{-0.039}$           \\
$w$                  & $-0.880^{+0.039}_{-0.040}$   & $-0.877 \pm 0.036$   \\
\midrule
$\sigma_{\rm host}$  & ---                           & $0.860^{+0.134}_{-0.111}$     \\
$e^\mu$              & ---                           & $95.0^{+16.3}_{-15.8}$         \\
\bottomrule
\end{tabular}
\caption{Marginalized constraints at 68\% confidence level for the $w$CDM model, obtained from 
the baseline dataset and from the combined dataset including FRBs.}
\label{tab:wcdm_constraints}
\end{table}

This behavior is illustrated in Fig.~\ref{fig:wcdm_corner}, where the inclusion of FRBs leads
to a visible contraction of the posterior contours, particularly along directions involving 
$\Omega_{\rm b}$. However, the degeneracies associated with the dark-energy parameter persist,
limiting the propagation of the FRB information to the expansion history. As a result, the 
additional freedom introduced by $w$ partially absorbs the constraining power of the FRB data,
reducing their impact on the dark-energy sector compared to the standard $\Lambda$CDM model.

\subsection{CPL}
\label{sec:results_cpl}

\begin{figure}[h!]
\centering
\includegraphics[width=0.85\linewidth]{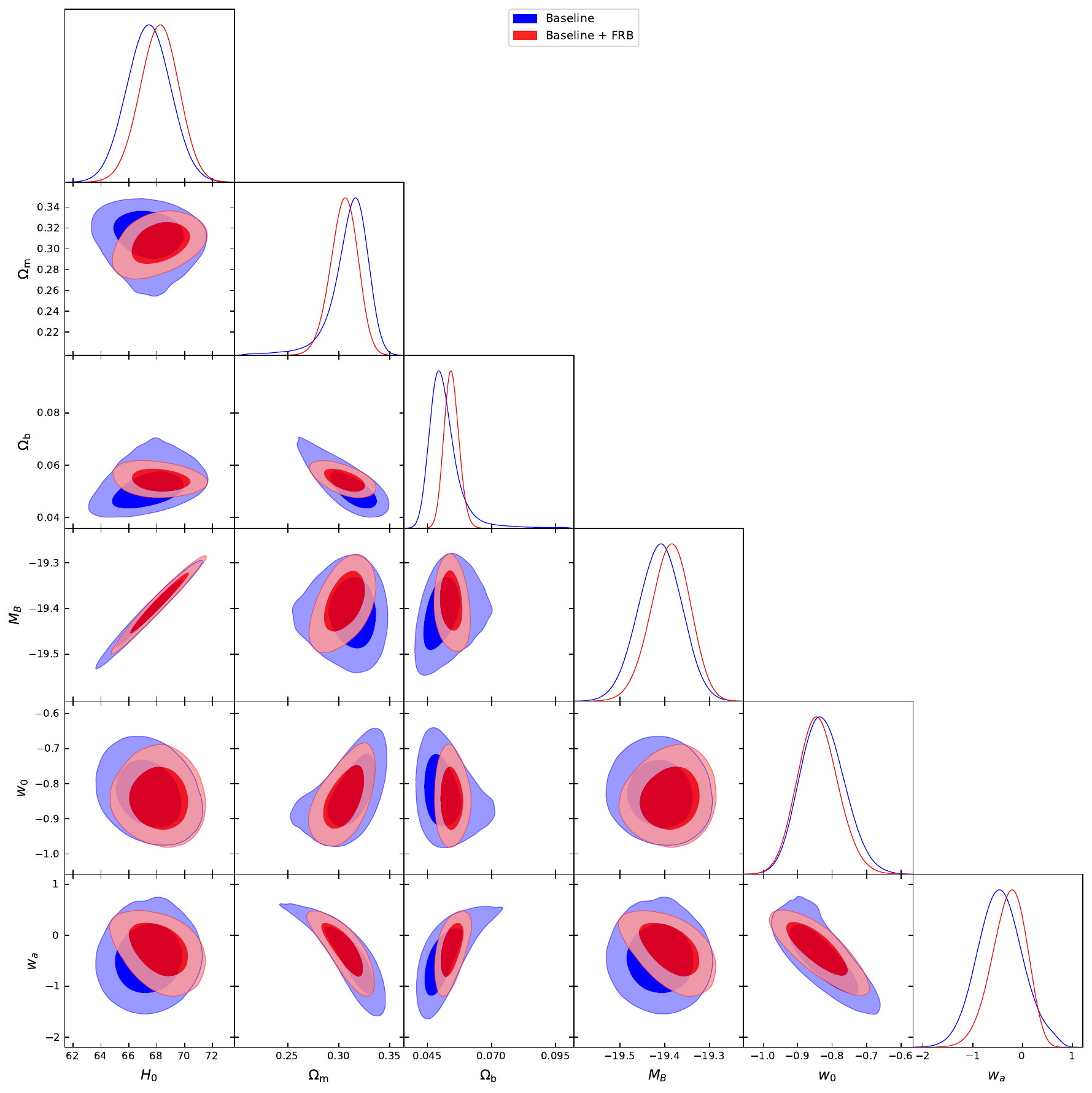}
\caption{Posterior distributions for the CPL model for the Baseline dataset (blue) and including
FRBs (red). Contours correspond to the 68\% and 95\% credible regions.}
\label{fig:cpl_corner}
\end{figure}

The posterior distributions for the CPL model are shown in Fig.~\ref{fig:cpl_corner}, with the 
corresponding marginalized constraints reported in Table~\ref{tab:cpl_constraints}.
The inclusion of FRB data improves the constraints on several cosmological parameters, although
the gains are generally smaller than those obtained in the simpler $\Lambda$CDM and $w$CDM scenarios. 
The largest improvement is observed in the baryon density $\Omega_{\rm b}$, whose uncertainty reduced
from $\sim 0.0051$ to $\sim 0.0029$ (a $43.1\%$ improvement). On the other hand, the uncertainty in 
$H_0$ decreased from $\sim 1.56$ to $\sim 1.38$ km\,s$^{-1}$\,Mpc$^{-1}$ (a $11.5\%$ improvement), 
and the uncertainty in $M_B$ decreased from $\sim 0.049$ to $\sim 0.044$ (a $10.2\%$ improvement). 
The dark energy sector, in turn, shows limited improvement in $w_0$, with the uncertainty decreasing 
from $\sim 0.064$ to $\sim 0.059$ (a $7.8\%$ improvement), while $w_a$ exhibits a moderate improvement,
with its uncertainty decreasing from $\sim 0.45$ to $\sim 0.35$ (a $22.2\%$ improvement).

\begin{table}
\centering
\begin{tabular}{lcc}
\toprule
\textbf{Parameter}   & \textbf{Baseline}      & \textbf{Baseline + FRB} \\
\midrule
$H_0$                & $67.42 \pm 1.56$             & $68.20^{+1.36}_{-1.40}$       \\
$\Omega_{\rm m}$     & $0.313^{+0.014}_{-0.018}$    & $0.306^{+0.013}_{-0.014}$     \\
$\Omega_{\rm b}$     & $0.0506^{+0.0059}_{-0.0042}$ & $0.0543^{+0.0030}_{-0.0028}$  \\
$M_B$                & $-19.410^{+0.048}_{-0.049}$  & $-19.386^{+0.043}_{-0.045}$   \\
$w_0$                & $-0.829^{+0.067}_{-0.061}$   & $-0.842^{+0.060}_{-0.057}$    \\
$w_a$                & $-0.457^{+0.457}_{-0.442}$   & $-0.261^{+0.327}_{-0.371}$    \\
\midrule
$\sigma_{\rm host}$  & ---                           & $0.851^{+0.132}_{-0.108}$    \\
$e^\mu$              & ---                           & $98.6^{+17.0}_{-16.7}$       \\
\bottomrule
\end{tabular}
\caption{Marginalized constraints at 68\% confidence level for the CPL model, obtained from
the baseline dataset and from the combined dataset including FRBs.}
\label{tab:cpl_constraints}
\end{table}

The reduced impact of FRBs in the CPL model originates from the enlarged parameter space
introduced by the $(w_0,w_a)$ sector. Strong correlations involving $H_0$, $w_0$, and $w_a$
absorb part of the constraining power provided by the FRB data, limiting the propagation of
the baryon-density information to the expansion history. This behavior is clearly illustrated
in Fig.~\ref{fig:cpl_corner}, where the posterior contours exhibit a visible contraction, 
particularly along directions associated with $\Omega_{\rm b}$, while the extended degeneracies
in the $(w_0,w_a)$ plane persist. Consequently, although FRBs substantially improve the 
determination of the baryon content, their ability to constrain the dark-energy sector remains
limited in the CPL parametrization.

\subsection{Appleby-Battye}
\label{sec:results_ab}

\begin{figure}
\centering
\includegraphics[width=0.85\linewidth]{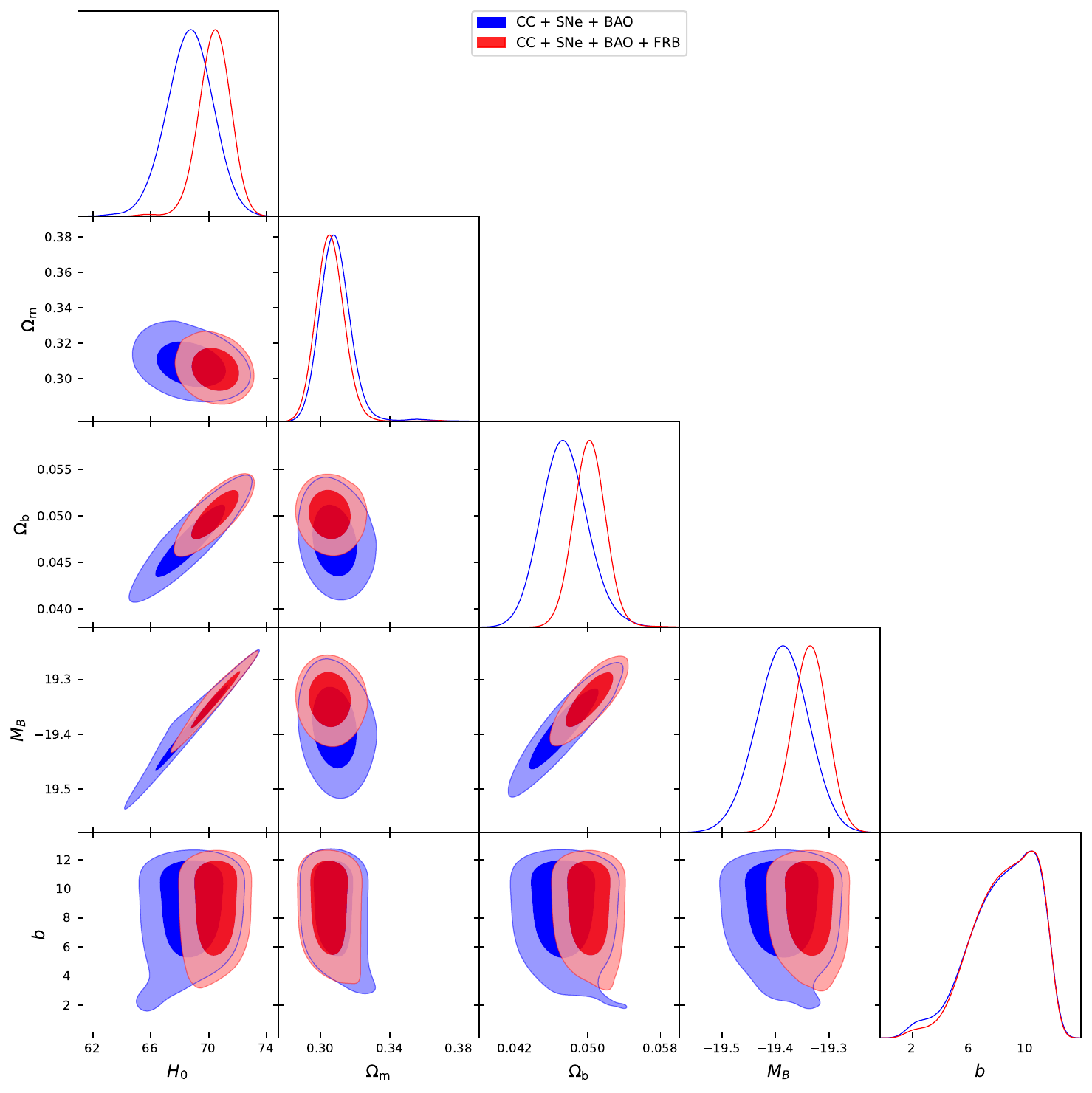}
\caption{Posterior distributions for the AB model for the Baseline dataset (blue) and 
including FRBs (red). Contours correspond to the 68\% and 95\% credible regions.}
\label{fig:ab_corner}
\end{figure}

The posterior distributions for the AB model are shown in Fig.~\ref{fig:ab_corner}, with
the corresponding marginalized constraints reported in Table~\ref{tab:ab_constraints}.
The inclusion of FRB data leads to substantial improvements in the cosmological parameter
constraints, closely resembling the behavior observed in the $\Lambda$CDM scenario. 
In particular, the uncertainty in the Hubble constant $H_0$ decreases from $\sim 1.59$ to
$\sim 1.08$ km,s$^{-1}$,Mpc$^{-1}$ (a $31.9\%$ improvement), while the uncertainty in the
baryon density $\Omega_{\rm b}$ decreases from $\sim 0.0025$ to $\sim 0.0017$ (a $32.0\%$ 
improvement). The uncertainty in the SNe~Ia absolute magnitude $M_B$ also decreases from 
$\sim 0.049$ to $\sim 0.033$ (a $33.0\%$ improvement).

This behavior is illustrated in Fig.~\ref{fig:ab_corner}, where the inclusion of FRB data 
produces a noticeable contraction and partial rotation of the posterior contours, particularly
in the $H_0$--$\Omega_{\rm b}$ and $H_0$--$M_B$ planes. The similarity between the AB and 
$\Lambda$CDM results reflects the fact that the AB expansion history rapidly approaches the 
standard cosmological model over a wide region of parameter space.

The additional parameter $b$ remains only weakly constrained, with the lower bound shifting 
slightly from $b > 6.123$ to $b > 6.319$ after the inclusion of FRBs. As $b$ increases, the
AB model rapidly approaches the $\Lambda$CDM limit, becoming effectively indistinguishable 
from it at the level of current observations. Consequently, the likelihood becomes increasingly
flat for $b \gtrsim 6$, leading to a posterior dominated by prior volume at large values of $b$.
As a result, neither a meaningful upper bound nor a well-defined central value can be determined, 
indicating that current late-time observations remain largely insensitive to deviations from 
$\Lambda$CDM within this class of models.

\begin{table}
\centering
\begin{tabular}{lcc}
\toprule
\textbf{Parameter}   & \textbf{Baseline}      & \textbf{Baseline + FRB} \\
\midrule
$H_0$                & $68.71^{+1.56}_{-1.61}$      & $70.42^{+1.06}_{-1.10}$       \\
$\Omega_{\rm m}$     & $0.308^{+0.009}_{-0.008}$    & $0.305 \pm 0.008$             \\
$\Omega_{\rm b}$     & $0.0474^{+0.0026}_{-0.0024}$ & $0.0502 \pm 0.0017$           \\
$M_B$                & $-19.387^{+0.048}_{-0.049}$  & $-19.336^{+0.032}_{-0.033}$   \\
$b$                  & $> 6.123$                    & $> 6.319$                     \\
\midrule
$\sigma_{\rm host}$  & ---                          & $0.834^{+0.123}_{-0.102}$     \\
$e^\mu$              & ---                          & $104.3^{+16.1}_{-15.2}$      \\
\bottomrule
\end{tabular}
\caption{Marginalized constraints at 68\% confidence level for the AB model, obtained from
the baseline dataset and from the combined dataset including FRBs.}
\label{tab:ab_constraints}
\end{table}

\subsection{Hu-Sawicki}
\label{sec:results_hs}

\begin{figure}[h!]
\centering
\includegraphics[width=0.85\linewidth]{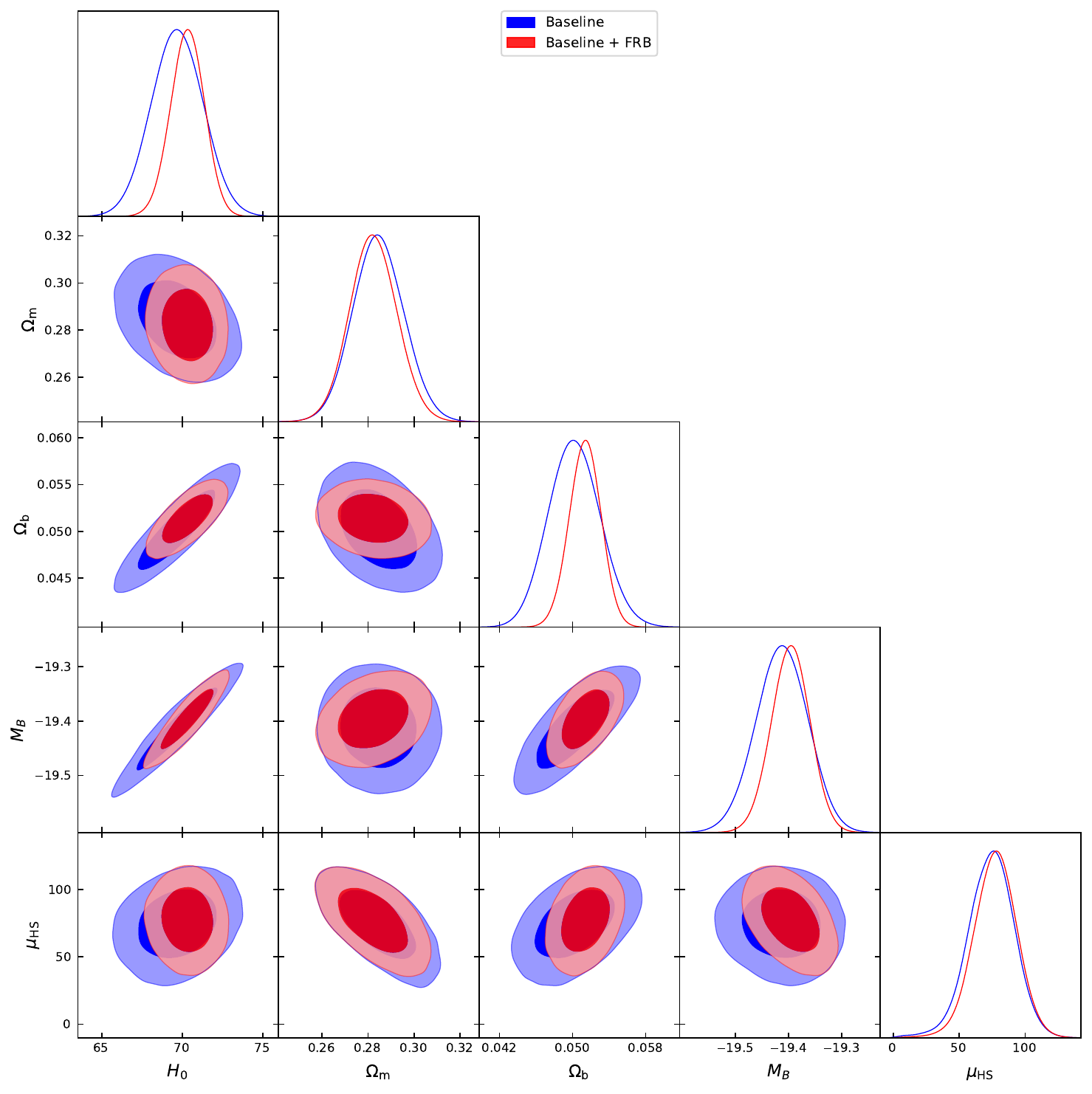}
\caption{Posterior distributions for the HS model for the Baseline dataset (blue) and 
including FRBs (red). Contours correspond to the 68\% and 95\% credible regions.}
\label{fig:hs_corner}
\end{figure}

The posterior distributions for the HS model ($n=1$) are shown in Fig.~\ref{fig:hs_corner}, 
with the corresponding marginalized constraints reported in Table~\ref{tab:hs_constraints}.
The inclusion of FRB data leads to substantial improvements in the cosmological parameter 
constraints, comparable to, or slightly stronger than, those obtained in the AB model. 
In particular, the uncertainty in the Hubble constant $H_0$ decreases from $\sim 1.60$ to
$\sim 1.04$ km,s$^{-1}$,Mpc$^{-1}$ (a $34.8\%$ improvement), while the uncertainty in the 
baryon density $\Omega_{\rm b}$ decreases from $\sim 0.0029$ to $\sim 0.0017$ (a $40.4\%$ 
improvement). The uncertainty in the SNe~Ia absolute magnitude $M_B$ also decreases from 
$\sim 0.049$ to $\sim 0.036$ (a $25.8\%$ improvement).

As illustrated in Fig.~\ref{fig:hs_corner}, the inclusion of FRB data produces a visible
contraction and mild rotation of the posterior contours, particularly in the 
$H_0$--$\Omega_{\rm b}$ and $H_0$--$M_B$ planes. The overall behavior closely resembles
that observed in the AB model, indicating that FRBs efficiently constrain the background 
cosmological parameters while the modified-gravity sector remains comparatively less affected.

\begin{table}
\centering
\begin{tabular}{lcc}
\toprule
\textbf{Parameter}  & \textbf{Baseline}      & \textbf{Baseline + FRB} \\
\midrule
$H_0$                & $69.69^{+1.60}_{-1.59}$      & $70.32^{+1.03}_{-1.05}$       \\
$\Omega_{\rm m}$     & $0.284 \pm 0.011$            & $0.282 \pm 0.010$             \\
$\Omega_{\rm b}$     & $0.0502^{+0.0029}_{-0.0028}$ & $0.0514 \pm 0.0017$           \\
$M_B$                & $-19.412^{+0.048}_{-0.049}$  & $-19.396 \pm 0.036$           \\
$\mu_{\rm HS}$       & $74.59^{+16.55}_{-17.45}$    & $77.54^{+15.68}_{-16.24}$     \\
\midrule
$\sigma_{\rm host}$  & ---                           & $0.857^{+0.132}_{-0.109}$    \\
$e^\mu$              & ---                           & $96.7^{+16.4}_{-15.9}$       \\
\bottomrule
\end{tabular}
\caption{Marginalized constraints at 68\% confidence level for the HS model, obtained from 
the baseline dataset and from the combined dataset including FRBs.}
\label{tab:hs_constraints}
\end{table}

The modified gravity parameter $\mu_{\rm HS}$ remains only weakly constrained, with a broad 
posterior distribution and no significant reduction in its uncertainty after the inclusion of 
FRBs. Although the posterior peaks at a non-zero value, corresponding to an approximately 
$4.7\sigma$ offset from the GR limit $\mu_{\rm HS}=0$, this should not be interpreted as evidence
for deviations from $\Lambda$CDM. Instead, it reflects the bounded and non-Gaussian nature of the 
posterior distribution, together with volume effects in parameter space. As shown in 
Fig.~\ref{fig:hs_corner}, $\mu_{\rm HS}$ exhibits moderate correlations with $H_0$, $\Omega_{\rm m}$,
and $\Omega_{\rm b}$, but these correlations are insufficient to significantly tighten its constraint.
Unlike the AB model, however, the posterior retains a well-defined peak and is not dominated by prior
volume, indicating that current late-time observations still provide limited, but non-negligible,
sensitivity to this parameter.

\subsubsection{Starobinsky}
\label{sec:results_st}

\begin{figure}[htbp]
\centering
\includegraphics[width=0.85\linewidth]{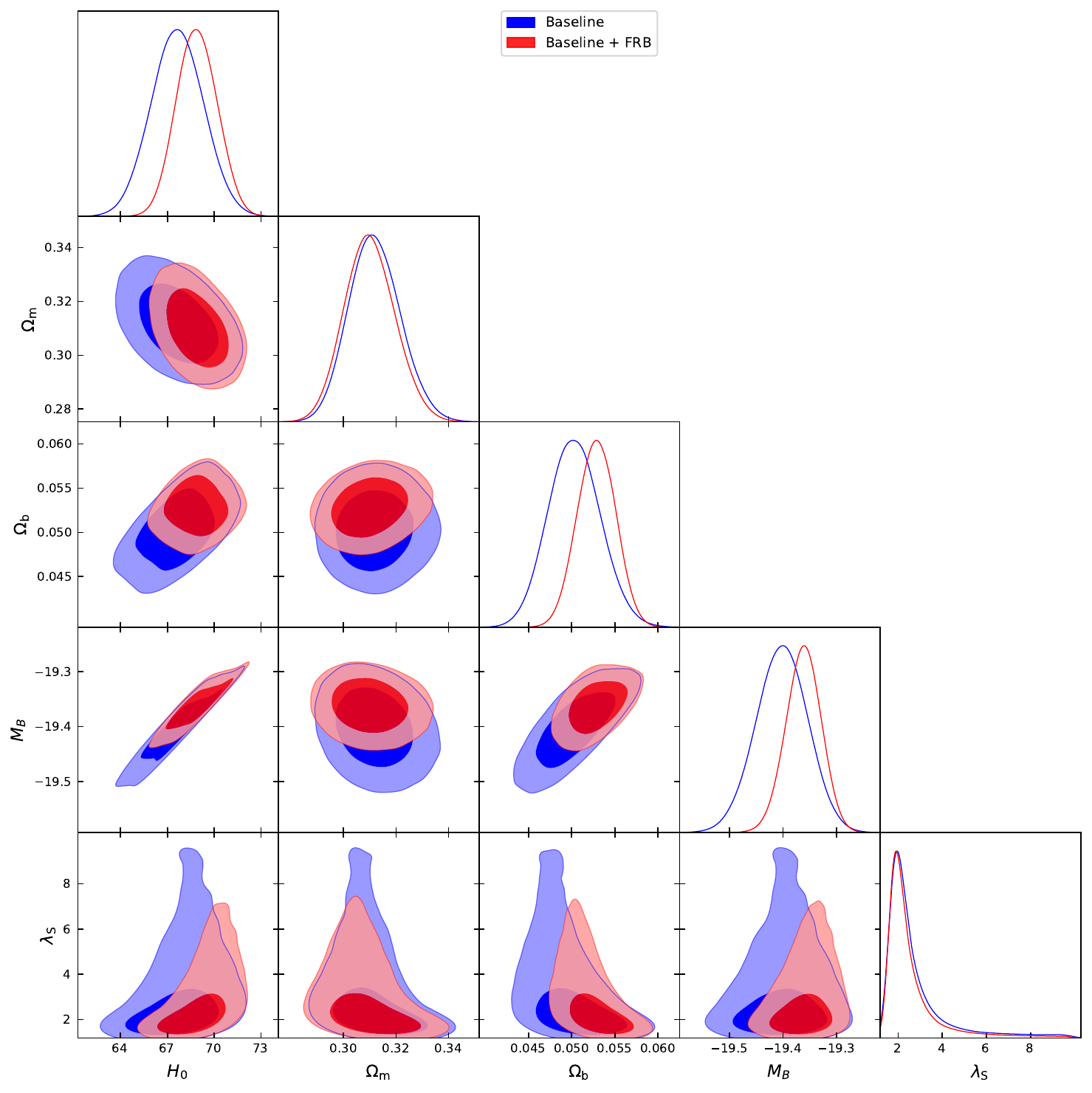}
\caption{Posterior distributions for the ST model for the baseline dataset (blue) and including FRBs
(red). Contours correspond to the 68\% and 95\% credible regions.}
\label{fig:st_corner}
\end{figure}

The posterior distributions for the ST model are shown in Fig.~\ref{fig:st_corner}, with the
corresponding marginalized constraints reported in Table~\ref{tab:st_constraints}.
The inclusion of FRB data improves the cosmological parameter constraints, although the gains
are generally smaller than those obtained in the AB and HS models. In particular, the uncertainty
in the Hubble constant $H_0$ decreases from $\sim 1.67$ to $\sim 1.30$ km,s$^{-1}$,Mpc$^{-1}$ 
(a $21.9\%$ improvement), while the uncertainty in the baryon density $\Omega_{\rm b}$ decreases 
from $\sim 0.0030$ to $\sim 0.0022$ (a $25.4\%$ improvement). A larger gain is observed for the 
SNe~Ia absolute magnitude $M_B$, whose uncertainty decreases from $\sim 0.048$ to $\sim 0.033$ 
(a $31.6\%$ improvement).

As illustrated in Fig.~\ref{fig:st_corner}, the inclusion of FRB data produces a visible contraction
and mild rotation of the posterior contours, particularly in the $H_0$--$M_B$ plane. The comparatively
larger improvement observed for $M_B$ reflects the partial alleviation of its degeneracy with the Hubble
constant, whereas the impact on the remaining cosmological parameters is more moderate than in the AB
and HS scenarios.

The modified-gravity parameter $\lambda_{\rm S}$ remains only weakly constrained, exhibiting a broad and
asymmetric posterior distribution with no significant reduction in its uncertainty after the inclusion 
of FRBs. Nevertheless, the posterior displays a well-defined peak around $\lambda_{\rm S}\sim 2$ and does 
not extend toward arbitrarily large values, indicating that the likelihood does not become fully flat in 
the $\Lambda$CDM limit. This behavior suggests that, unlike the AB model, current observations retain some
sensitivity to the modified-gravity sector, although the overall constraining power remains limited.

\begin{table}[htbp]
\centering
\begin{tabular}{lcc}
\toprule
\textbf{Parameter}  & \textbf{CC + SNe + BAO}       & \textbf{CC + SNe + BAO + FRB} \\
\midrule
$H_0$                & $67.64^{+1.67}_{-1.66}$      & $68.87^{+1.32}_{-1.28}$       \\
$\Omega_{\rm m}$     & $0.312^{+0.010}_{-0.009}$    & $0.310^{+0.010}_{-0.009}$     \\
$\Omega_{\rm b}$     & $0.0503^{+0.0030}_{-0.0029}$ & $0.0529 \pm 0.0022$           \\
$M_B$                & $-19.402^{+0.047}_{-0.048}$  & $-19.362^{+0.032}_{-0.033}$   \\
$\lambda_{\rm S}$    & $2.17^{+1.29}_{-0.47}$       & $2.10^{+1.08}_{-0.42}$        \\
\midrule
$\sigma_{\rm host}$  & ---                           & $0.842^{+0.127}_{-0.105}$    \\
$e^\mu$              & ---                           & $101.2^{+16.1}_{-15.7}$             \\
\bottomrule
\end{tabular}
\caption{Marginalized constraints at 68\% confidence level for the ST model, obtained from the 
baseline dataset and from the combined dataset including FRBs.}
\label{tab:st_constraints}
\end{table}

\subsection{FRB nuisance parameters}
\label{sec:results_frb_nuisance}

The posterior distributions for the FRB nuisance parameters, $\sigma_{\rm host}$
and $e^\mu$, exhibit reasonable stability across the diverse set of cosmological
scenarios explored in this work, as evidenced by the marginalized constraints 
reported in Tables~\ref{tab:lcdm_constraints}--\ref{tab:st_constraints}. For the
intrinsic host scatter, we find consistent constraints within the interval 
$\sigma_{\rm host} \in [0.833, 0.860]$, with relative $1\sigma$ uncertainties 
ranging from approximately $12\%$ to $16\%$. Similarly, the log-normal normalization
parameter $e^\mu$ remains well constrained within the range $95.0$--$104.3$, 
exhibiting relative $1\sigma$ uncertainties of about $15\%$--$17\%$ across all
gravitational frameworks.

We further emphasize that even in the presence of additional degrees of freedom,
such as the models analyzed here, the posterior distributions for the nuisance 
sector do not exhibit significant broadening or signs of prior dominance. The 
stability of the relative uncertainties suggests that the current FRB dataset 
possesses sufficient sensitivity to mitigate potential degeneracies between 
astrophysical host contributions and the underlying cosmological signal.
Consequently, the robustness of these constraints across both GR-based and 
modified gravity frameworks provides strong evidence for the internal consistency
of the FRB likelihood.

\subsection{The Impact of FRBs}
\label{sec:results_frb_impact}

We quantify the additional constraining power provided by FRB observations by 
evaluating both FoM and the $68.3\%$ credible-region area $A_{68.3}$ in the 
$(H_0,\Omega_{\rm b})$ plane, following the methodology described in Sec.~\ref{sec:fom}.
The complementarity between the Baseline dataset and FRBs is particularly important 
because, while the BAO scale depends on the baryon density through the sound horizon
at the drag epoch $r_d$, the dispersion measures of FRBs provide an independent 
late-time probe of the cosmic baryon content. This additional information helps to 
break the degeneracy between the expansion rate and the baryon density.

As summarized in Table~\ref{tab:fom_results}, the inclusion of FRB data systematically
reduces the area of the joint confidence regions and consequently increases the FoM for 
all cosmological scenarios considered. The reduction in $A_{68.3}$ ranges from 
approximately $33\%$ in the $\Lambda$CDM model to nearly $48\%$ in the CPL parametrization,
indicating a substantial contraction of the allowed parameter space. Correspondingly, the 
FoM increases by $48.4\%$ in the standard $\Lambda$CDM scenario and reaches a maximum
improvement of $90.8\%$ for the CPL model.
\begin{table}[h!]
\small
\centering
\begin{tabular}{lccccc}
\toprule
\multirow{2}{*}{\textbf{Model}}
& \multicolumn{2}{c}{\boldmath$A_{68.3}$}
& \multicolumn{2}{c}{\textbf{FoM}}
& \multirow{2}{*}{\textbf{Improvement}} \\
\cmidrule(lr){2-3}
\cmidrule(lr){4-5}
& \textbf{Baseline} & \textbf{+FRB}
& \textbf{Baseline} & \textbf{+FRB}
& \\
\midrule
$\Lambda$CDM & 0.0120 & 0.0081 & 83.6 & 124.0 & 48.4\% \\
$w$CDM       & 0.0455 & 0.0249 & 22.0 & 40.2  & 82.6\% \\
CPL          & 0.0564 & 0.0296 & 17.7 & 33.8  & 90.8\% \\
AB           & 0.0139 & 0.0090 & 71.7 & 111.2 & 55.1\% \\
HS           & 0.0157 & 0.0098 & 63.5 & 102.2 & 61.0\% \\
ST           & 0.0336 & 0.0215 & 29.8 & 46.6  & 56.4\% \\
\bottomrule
\end{tabular}
\caption{Comparison of the $68.3\%$ credible-region area and the Figure of Merit
in the $(H_0,\Omega_{\rm b})$ plane before and after the inclusion of FRB data.
The last column gives the relative improvement in the FoM due to the FRB contribution.}
\label{tab:fom_results}
\end{table}
\begin{table}[h!]
\small
\centering
\begin{tabular}{lcccccccc}
\toprule
\multirow{2}{*}{\textbf{Model}}
& \multicolumn{2}{c}{\boldmath$\sigma(H_0)$}
& \multicolumn{2}{c}{\boldmath$\sigma(\Omega_{\rm b})$}
& \multicolumn{2}{c}{\textbf{Skew}\boldmath$(H_0)$}
& \multicolumn{2}{c}{\textbf{Skew}\boldmath$(\Omega_{\rm b})$} \\
\cmidrule(lr){2-3}
\cmidrule(lr){4-5}
\cmidrule(lr){6-7}
\cmidrule(lr){8-9}
& \textbf{Baseline} & \textbf{+FRB}
& \textbf{Baseline} & \textbf{+FRB}
& \textbf{Baseline} & \textbf{+FRB}
& \textbf{Baseline} & \textbf{+FRB} \\
\midrule
$\Lambda$CDM & 1.524 & 1.048 & 0.00246 & 0.00168 & 0.000  & $-0.074$ & 0.121 & $-0.043$ \\
$w$CDM       & 1.557 & 1.298 & 0.00415 & 0.00253 & 0.023  & $-0.009$ & 0.314 & 0.089 \\
CPL          & 1.563 & 1.384 & 0.00504 & 0.00288 & 0.019  & $-0.077$ & 2.391 & 0.197 \\
AB           & 1.584 & 1.079 & 0.00253 & 0.00173 & $-0.226$ & $-0.601$ & 0.251 & 0.286 \\
HS           & 1.600 & 1.037 & 0.00281 & 0.00171 & 0.022  & $-0.065$ & 0.120 & $-0.046$ \\
ST           & 1.667 & 1.301 & 0.00299 & 0.00221 & 0.023  & 0.037 & 0.121 & $-0.023$ \\
\bottomrule
\end{tabular}
\caption{Marginal uncertainties and skewness coefficients of the one-dimensional posterior
distributions for $H_0$ and $\Omega_{\rm b}$ before and after the inclusion of FRB data.}
\label{tab:marginal_skew}
\end{table}

The largest gains are observed in the dynamical dark-energy scenarios, particularly CPL 
and $w$CDM, where parameter degeneracies are intrinsically stronger. In these cases, FRBs 
provide an additional constraint direction that efficiently reduces the volume of the 
posterior distribution. Similarly, the modified-gravity models AB, HS, and ST also exhibit
significant improvements, with FoM increases ranging from approximately $55\%$ to $61\%$. 
These results further demonstrate the enhanced constraining power of FRBs in extended 
cosmological scenarios, where additional degrees of freedom typically introduce stronger 
parameter degeneracies.

The impact of FRBs on the one-dimensional posterior distributions is summarized in 
Table~\ref{tab:marginal_skew}. The inclusion of FRB observations systematically reduces 
the marginal uncertainties of both $H_0$ and $\Omega_{\rm b}$, leading to more precise
parameter estimates across all cosmological models. In addition, the skewness coefficients
reveal that FRB data generally produce more symmetric posterior distributions. This effect
is particularly evident for the CPL model, where the skewness of the $\Omega_{\rm b}$ 
posterior decreases dramatically from $2.391$ to $0.197$, indicating a substantial reduction
in non-Gaussian features. Similar, although less pronounced, changes are observed in the
remaining models. Overall, FRBs not only tighten the cosmological constraints but also
regularize the posterior distributions, yielding likelihood surfaces that are closer to
Gaussian and consequently more robust for parameter inference.

\subsection{Model Comparison}
\label{sec:results_comparison}

To evaluate the statistical performance of the models considered in this work, we employ the
AIC, BIC, and LRT $p$-value, following the methodology outlined in Sec.~\ref{sec:selection}.
Table~\ref{tab:model_comparison} summarizes the statistical performance of each cosmological
framework against the full Baseline + FRB dataset, using $\Lambda$CDM as reference.

\begin{table}[h!]
\footnotesize
\centering
\begin{tabular}{lcccccc}
\toprule
\textbf{Quantity} & \textbf{$\Lambda$CDM} & \textbf{$w$CDM} & \textbf{CPL}
& \textbf{AB} & \textbf{HS} & \textbf{ST} \\
\midrule
$k$                     & 6            & 7            & 8            & 7            & 7            & 7            \\
$\ln\mathcal{L}_{\max}$ & $-1383.3235$ & $-1378.0069$ & $-1377.7934$ & $-1383.4515$ & $-1378.2402$ & $-1378.2307$ \\
$\chi^2$                & 2766.6470    & 2756.0138    & 2755.5868    & 2766.9030    & 2756.4804    & 2756.4614    \\
$\chi^2/\mathrm{dof}$   & 1.5955       & 1.5903       & 1.5910       & 1.5966       & 1.5906       & 1.5906       \\
AIC                     & 2778.6470    & 2770.0137    & 2771.5868    & 2780.9029    & 2770.4804    & 2770.4615    \\
BIC                     & 2811.4168    & 2808.2452    & 2815.2799    & 2819.1344    & 2808.7119    & 2808.6930    \\
\midrule
$\Delta\mathrm{AIC}$    & 0            & $-8.6333$    & $-7.0602$    & $+2.2559$    & $-8.1666$    & $-8.1855$    \\
$\Delta\mathrm{BIC}$    & 0            & $-3.1716$    & $+3.8631$    & $+7.7176$    & $-2.7049$    & $-2.7238$    \\
AIC verdict             & Ref.         & Str. alt.    & Str. alt.    & Pos. ref.    & Str. alt.    & Str. alt.    \\ 
BIC verdict             & Ref.         & Pos. alt.    & Pos. ref.    & Str. ref.    & Pos. alt.    & Pos. alt.    \\
LRT $p$-value           & ---          & 0.0011       & 0.0040       & 1.0000       & 0.0014       & 0.0014       \\
\bottomrule
\end{tabular}
\caption{Model comparison statistics for the full CC~+~SNe~+~BAO~+~FRB data combination,
using $\Lambda$CDM as the reference model. The abbreviations mean Pos. alt. (positive evidence 
for the alternative model); Pos. ref. (positive evidence for the reference model); Str. alt.
(strong evidence for the alternative model); and Str. ref. (strong evidence for the reference
model).}
\label{tab:model_comparison}
\end{table}

All extended cosmological scenarios considered in this work, except the AB model, provide an improved
fit relative to $\Lambda$CDM. In particular, the $w$CDM, CPL, HS, and ST frameworks reduce the best-fit 
$\chi^2$ by $|\Delta\chi^2| \approx 10$, corresponding to a modest decrease in $\chi^2/\mathrm{dof}$ 
from 1.596 to $\sim 1.591$. In contrast, the AB model yields a slightly worse fit, with $\Delta\chi^2 = +0.26$.

According to the AIC, the $w$CDM, CPL, HS, and ST models are strongly favored over $\Lambda$CDM, with
$\Delta\mathrm{AIC} < -7$, indicating that the improvement in fit outweighs the penalty associated with
the additional degrees of freedom. The AB model is instead mildly disfavored ($\Delta\mathrm{AIC} > +2$)
relative to the standard model.

When adopting the more conservative BIC, which imposes a stronger penalty on model complexity
($k \ln N \approx 7.46,k$), only $w$CDM, HS, and ST retain positive evidence relative to 
$\Lambda$CDM, with $-3.2 < \Delta\mathrm{BIC} < -2.7$. The CPL parametrization is penalized due
to its larger parameter space ($\Delta\mathrm{BIC} \simeq +3.9$), while the AB model is strongly
disfavored ($\Delta\mathrm{BIC} \simeq +7.7$).

The LRT further quantifies the statistical improvement achieved by nested extensions of the
$\Lambda$CDM model. The $w$CDM, CPL, HS, and ST models yield $p$-values ranging from $0.0011$ to
$0.0040$, indicating a statistically significant improvement over $\Lambda$CDM at the $\alpha = 0.05$ 
level, corresponding to an approximate $3\sigma$ preference for these extensions. In contrast, the AB
model yields $p = 1.000$, indicating no statistical preference relative to $\Lambda$CDM.

Overall, the results point to a moderate statistical preference for extensions beyond $\Lambda$CDM,
although the evidence remains inconclusive. While several models yield improved fits to the data, the 
corresponding BIC differences are limited to $|\Delta\mathrm{BIC}| < 3.2$, placing them within the
weak-to-positive evidence regime. This indicates that the observed gains in goodness of fit are not 
sufficient to decisively favor more complex models once the penalty for additional parameters is taken 
into account. These findings underscore the sensitivity of model selection to the trade-off between 
goodness of fit and model complexity, as well as to the choice of statistical criterion.


\section{Conclusion}
\label{sec:conclusion}

In this work, we have performed a comprehensive cosmological analysis combining CC, SNe~Ia,
BAO, and FRB datasets to constrain six cosmological scenarios: the standard $\Lambda$CDM
model, two dark energy parametrizations ($w$CDM and CPL), and three viable $f(R)$ models
(AB, HS, and ST). Our analyses employed MCMC methods to derive parameter constraints, 
quantified the impact of FRBs through the FoM in the $(H_0, \Omega_{\rm b})$ plane, and 
assessed model performance using AIC, BIC, and LRT statistics.

The inclusion of FRB data leads to significant improvements in cosmological constraints 
across all models. The most pronounced enhancements are observed in the baryon density
$\Omega_{\rm b}$, with relative uncertainty reductions ranging from $\sim 25\%$ to 
$\sim 43\%$, with the largest gain in the CPL parametrization and the smallest in the 
ST model, confirming the nearly model-independent constraining power of the DM--$z$ 
relation. For the Hubble constant $H_0$, the improvement is more model-dependent, 
spanning $\sim 12\%$ to $\sim 35\%$, with the weakest gains in CPL and the strongest 
in viable $f(R)$ scenarios such as HS, reflecting the role of residual degeneracies. 
A similar pattern is observed for the absolute magnitude $M_B$, with reductions between 
$\sim 10\%$ and $\sim 32\%$, again showing diminished impact in CPL and stronger gains 
in models with tighter parameter correlations. Overall, these results highlight that FRBs
provide a robust constraint on $\Omega_{\rm b}$, while their impact on other parameters 
is mediated by the structure of model-dependent degeneracies.

Another common feature observed across all cosmological scenarios is a systematic upward 
shift in the mean values of both $H_0$ and $\Omega_{\rm b}$ after the inclusion of FRB 
data. Although these shifts remain statistically consistent within the corresponding 
$1\sigma$ confidence intervals, they indicate that the FRB likelihood generally favors
slightly larger expansion rates and baryon fractions. The effect is particularly pronounced 
in the $\Lambda$CDM and AB models, where the mean value of $H_0$ increases by
$\sim 1.7$ km\,s$^{-1}$\,Mpc$^{-1}$, while $\Omega_b$ increases by $\sim0.003$.

In contrast, the matter density parameter $\Omega_{\rm m}$ remains remarkably stable 
across most cosmological scenarios. Since its constraints are already dominated by 
geometric probes such as CC, SNe, and BAO observations, the additional information 
provided by FRBs generally has only a minor impact on its posterior distribution. 
A modest improvement is observed only in the CPL model, where the larger parameter 
space allows part of the FRB information to propagate into $\Omega_{\rm m}$.

For the additional parameters in the GR-based extensions, the inclusion of FRBs 
yields more limited benefits. In the $w$CDM case, $w$ improves by $\sim 9\%$ and
remains consistent with $w = -1$. In the CPL parametrization, $(w_0, w_a)$ show 
uneven gains ($\sim 8\%$ and $\sim 22\%$, respectively), with $w_a$ still weakly
constrained. Although the marginal constraint on $w_0$ places $w_0=-1$ outside 
the $68\%$ confidence interval, the $\Lambda$CDM limit is not significantly 
disfavored, since $w_a=0$ remains compatible with the data and the two-dimensional 
$(w_0,w_a)$ constraints are considerably broader than the marginalized intervals. 
A qualitatively similar behavior has been reported in some DESI data 
combinations~\cite{Abdul_Karim_2025}.

Similarly, in $f(R)$ models, only marginal improvements are obtained: only a 
lower bound on $b$ could be established, while the uncertainties on $\mu_{\rm HS}$
and $\lambda_{\rm S}$ decreased by $\sim 6\%$ and $\sim 15\%$, respectively, 
although the latter remains characterized by highly asymmetric uncertainties.
No significant shifts were found in the central values of these parameters.
Overall, these results indicate that FRBs have a limited impact on the additional 
degrees of freedom, with the constraints remaining largely dominated by parameter
degeneracies.

The FRB nuisance parameters $\sigma_{\rm host}$ and $e^\mu$ exhibit only moderate
variations across the different cosmological scenarios, with their inferred values
remaining within relatively narrow ranges. This behavior suggests that these
quantities are primarily determined by astrophysical properties of the host
environment, with only a limited dependence on the underlying cosmological model.
Consequently, the observed improvements in cosmological constraints appear to be
driven mainly by the FRB cosmological information rather than by substantial shifts 
in the nuisance parameters.

The global impact of FRBs is quantified by the HPD-based FoM in the 
$(H_0, \Omega_{\rm b})$ plane, which increases by $\sim 48\%$ in $\Lambda$CDM case,
$\sim 83\%$ in $w$CDM, and up to $\sim 91\%$ in CPL parametrizarion. This substantial
enhancement is driven by the orthogonal constraint provided by the DM--$z$ relation, 
which directly probes the integrated electron density and effectively breaks 
degeneracies between the expansion rate and the baryon content. Furthermore, FRBs
induce a systematic reduction in parameter correlations and posterior skewness, 
leading to more Gaussian likelihood surfaces and more robust cosmological inferences.

The model comparison analysis reveals a moderate statistical preference for extensions
beyond $\Lambda$CDM. Four alternative models, $w$CDM, CPL, HS, and ST, achieve improved
fits to the data, reducing $\chi^2$ by $|\Delta\chi^2| \approx 10$ relative to $\Lambda$CDM. 
Under the AIC, these models show strong evidence ($\Delta\mathrm{AIC} < -7$), indicating 
that the improved fits compensate for the additional model complexity. However, the more 
conservative BIC penalizes the CPL model due to its larger parameter space, retaining 
positive evidence only for $w$CDM, HS, and ST ($-3.2 < \Delta\mathrm{BIC}| < -2.7 $). 
The LRT $p$-values, in turn, indicate an improvement over $\Lambda$CDM, with 
$p \sim 0.001$--$0.004$, corresponding to approximately $3\sigma$ statistical significance.
However, this evidence should be interpreted with caution, as it remains well below the
$5\sigma$ threshold typically required for a robust detection.

The comparable performance of $w$CDM and $f(R)$ models underscores a fundamental degeneracy. 
At the level of background observables, dark energy and modified gravity can produce nearly 
indistinguishable expansion histories. Since the probes employed in this work are primarily
sensitive to $H(z)$ and geometric distances, they cannot decisively discriminate between 
these frameworks. Breaking this degeneracy requires the inclusion of perturbation-level 
observables that probe the growth of cosmic structure.

Future progress will require both larger datasets and complementary observables sensitive
to structure formation. In particular, redshift-space distortions, weak gravitational
lensing, the integrated Sachs-Wolfe effect, and cluster abundances provide information on
the growth rate $f\sigma_8(z)$ and the metric potentials, which differ between dark energy 
and modified gravity even when their expansion histories are identical. The combination of 
these probes with an expanded FRB catalog will enable joint analyses that simultaneously 
constrain the expansion history, the baryon content, and the growth of structure.

For FRBs specifically, improvements in systematic control are essential. Current 
uncertainties are dominated by the modeling of the intergalactic medium ionization state, 
host galaxy contributions to the dispersion measure, and the precision of redshift
measurements. As localization techniques improve and larger samples of FRBs with confirmed
host galaxy associations become available, these systematic uncertainties are expected to 
decrease substantially. Refined theoretical models of the baryon distribution and improved 
calibration of the DM--$z$ relation will enhance the cosmological constraining power of 
FRBs, especially at intermediate redshifts ($z \sim 0.5$--$1.5$), where current data are sparse.

Finally, this work demonstrates that FRBs may be a powerful and complementary probe of the 
cosmic baryon distribution, capable of significantly enhancing the constraining power of 
late-time cosmological datasets. While current data favor extensions beyond $\Lambda$CDM 
with moderate statistical significance, decisive discrimination between competing cosmological 
scenarios will require the combination of larger samples, improved systematic control, and the
inclusion of observables sensitive to both background and perturbation dynamics.


\acknowledgments

B.W.N.R., L.L.S., K.E.L.F., R.A.B., and A.R.Q. thank the Para\'iba State Research Foundation 
(FAPESQ) for financial support.  A.R.Q. also acknowledges the support by CNPq under process 
number 310533/2022-8. K.E.L.F. and R.H.S. also thank CNPq for financial support.


\appendix

\section{\texorpdfstring{$f(R)$ Background Dynamics}{f(R) Background Dynamics}}
\label{app:ode}

In this appendix we outline the procedure used to determine the background expansion
history $H(z)$ in $f(R)$ cosmology. The modified Friedmann equations~(\ref{eq:fr_friedmann1})
and~(\ref{eq:fr_friedmann2}) are recast into differential systems suitable for numerical 
integration, with formulations adapted to the specific structure of each model. In all cases,
the ODE system is integrated forward in the scale factor from $a_i = 0.2$ to $a_f = 1.0$, 
corresponding to the redshift range $z \in [0, 4]$; the resulting solution is then mapped 
to $H(z)$ via the standard relation $a = (1+z)^{-1}$. The initialization at $a_i$ is motivated 
by the fact that, for $z \gtrsim 4$, deviations from GR are strongly suppressed, so the
background dynamics can be accurately approximated by the $\Lambda$CDM solution. The $H(z)$ 
profiles obtained in this way are subsequently used to construct a dense interpolation grid
for efficient likelihood evaluations within the MCMC sampler.

\paragraph{Appleby-Battye model.}
For the AB model, the background dynamics can be expressed as a third-order ODE for the
Hubble parameter as a function of the scale factor, $H(a)$. Adopting $a$ as the independent
variable and rewriting time derivatives via $\frac{d}{dt} = aH\frac{d}{da}$, the Ricci 
scalar takes the form
\begin{equation}
    R = 6H\left(aH^\prime + 2H\right),
\end{equation}
where primes denote derivatives with respect to $a$. Substituting this relation into 
the modified Friedmann equations~(\ref{eq:fr_friedmann1}) and~(\ref{eq:fr_friedmann2}),
the system reduces to
\begin{equation}
    H^{\prime\prime\prime}(a) = \mathcal{F}\left(a, H, H^\prime, H^{\prime\prime}\right),
    \label{eq:ab_ode}
\end{equation}
where $\mathcal{F}$ encodes the full non-linear dependence on $f(R)$ and its derivatives. 
The third-order nature of this equation requires three initial conditions, set consistently
with the $\Lambda$CDM background at $a_i$:
\begin{equation}
    H_i \equiv H_{\Lambda \rm CDM}(a_i)\,,\quad H^\prime_i \equiv H^\prime_{\Lambda \rm CDM}(a_i)\,,\quad H^{\prime\prime}_i \equiv H^{\prime\prime}_{\Lambda \rm CDM}(a_i)\,.
\end{equation}

\paragraph{Hu-Sawicki and Starobinsky models.}
For the HS and ST models, we follow the strategy introduced in~\cite{Hu_2007}, which
reformulate the background evolution in terms of the dimensionless variable $y_H(N)$,
defined as $y_H \equiv \left(H^2/m^2\right) - a^{-3}$, where $N \equiv \ln a$ 
and $m^2 = H_0^2 \Omega_{\rm m}$.

\paragraph{$R^2$ regularization.}
All three models include an $R^2$ regularization term whose role is to suppress
high-frequency oscillations of the Ricci scalar and prevent the emergence of 
curvature singularities, while preserving the correct late-time cosmological 
behavior. The frequency of these small oscillations, 
$\omega_{\rm osc} \equiv M_{\rm s} \leq M$, depends on $R$ and is therefore 
time-dependent. It is upper-bounded by the mass scale $M$, which is very large 
during inflation ($\sim 10^{13}\,\text{GeV}$) and decreases as the universe ages.
Following~\cite{Appleby_2010}, an appropriate mass scale corresponds to 
$\delta_s \equiv \epsilon_{f(R)}/M^2 = 10^{-7}$ for the redshift range of interest 
($0 \leq z \leq 4$). The characteristic curvature scale $\epsilon_{f(R)}$ takes
model-specific values: $\epsilon_{\rm AB}$ for the AB model, $m^2$ for the HS model, 
and $R_{\rm S}$ for the ST model.

\paragraph{Code availability.}
The numerical routines used to integrate the background equations for all models,
along with the complete MCMC pipeline, are publicly available at \url{https://github.com/bwesley92/MCMC-cosmo}.


\section{CC and FRB Catalogs}

This appendix lists the CC and FRB datasets described in
Secs.~\ref{sec:cc} and~\ref{sec:data_frb}, provided for completeness
and reproducibility. The DESI DR2 BAO and Pantheon$+$ SNe~Ia data are
not reproduced here as they are publicly available from their respective
collaborations.

\begin{table}
\centering
\begin{tabular}{S S S c | S S S c}
\toprule
{$z$} & {$H(z)$} & {$\sigma_{H}(z)$} & {Ref.} & {$z$} & {$H(z)$} & {$\sigma_{H}(z)$} & {Ref.} \\
\midrule
0.07   & 69.0 & 19.6 & \cite{Zhang_2014}   & 0.593 & 104.0 & 13.0 & \cite{Moresco_2012}   \\
0.09   & 69.0 & 12.0 & \cite{Simon_2005}   & 0.68  & 92.0  & 8.0  & \cite{Moresco_2012}   \\
0.12   & 68.6 & 26.2 & \cite{Zhang_2014}   & 0.75  & 98.8  & 33.6 & \cite{Borghi_2022}    \\
0.17   & 83.0 & 8.0  & \cite{Simon_2005}   & 0.75  & 105.0 & 7.9  & \cite{Jimenez_2023}   \\
0.179  & 75.0 & 4.0  & \cite{Moresco_2012} & 0.781 & 105.0 & 12.0 & \cite{Moresco_2012}   \\
0.199  & 75.0 & 5.0  & \cite{Moresco_2012} & 0.8   & 113.1 & 15.1 & \cite{Jiao_2023}      \\
0.2    & 72.9 & 29.6 & \cite{Zhang_2014}   & 0.875 & 125.0 & 17.0 & \cite{Moresco_2012}   \\
0.27   & 77.0 & 14.0 & \cite{Simon_2005}   & 0.88  & 90.0  & 40.0 & \cite{Stern_2010}     \\
0.28   & 88.8 & 36.6 & \cite{Zhang_2014}   & 0.9   & 117.0 & 23.0 & \cite{Simon_2005}     \\
0.352  & 83.0 & 14.0 & \cite{Moresco_2012} & 1.037 & 154.0 & 20.0 & \cite{Moresco_2012}   \\
0.3802 & 83.0 & 13.5 & \cite{Moresco_2016} & 1.26  & 135.0 & 65.0 & \cite{Tomasetti_2023} \\
0.4    & 95.0 & 17.0 & \cite{Simon_2005}   & 1.3   & 168.0 & 17.0 & \cite{Simon_2005}     \\
0.4004 & 77.0 & 10.2 & \cite{Moresco_2016} & 1.363 & 160.0 & 33.6 & \cite{Moresco_2015}   \\
0.4247 & 87.1 & 11.2 & \cite{Moresco_2016} & 1.43  & 177.0 & 18.0 & \cite{Simon_2005}     \\
0.4497 & 92.8 & 12.9 & \cite{Moresco_2016} & 1.53  & 140.0 & 14.0 & \cite{Simon_2005}     \\
0.47   & 89.0 & 49.6 & \cite{Rats_2017}    & 1.75  & 202.0 & 40.0 & \cite{Simon_2005}     \\
0.4783 & 80.9 & 9.0  & \cite{Moresco_2016} & 1.965 & 186.5 & 50.4 & \cite{Moresco_2015}   \\
0.48   & 97.0 & 62.0 & \cite{Stern_2010}   &       &       &      &                      \\
\bottomrule
\end{tabular}
\caption{Catalog of 35 $H(z)$ measurements from CC, spanning $0.07 \leq z \leq 1.965$,
compiled from multiple observational studies. Values are expressed in units of 
km\,s$^{-1}$\,Mpc$^{-1}$ with $1\sigma$ uncertainties.}
\label{tab:cc}
\end{table}

\setlength{\LTcapwidth}{\textwidth}
\begin{longtable}{l S S S c c}
\caption{Catalog of 104 FRBs with measured $\mathrm{DM_{obs}}$,
redshifts, and identified host galaxies, covering the redshift
range $0.0085 \leq z \leq 1.354$, compiled from multiple surveys
and follow-up observations.}\\
\toprule
{Name} & {$z$} & {$\mathrm{DM_{obs}}(z)$} & {$\mathrm{DM_{MW,ISM}}(z)$} & Host & Ref. \\
\midrule
\endfirsthead
\toprule
{Name} & {$z$} & {$\mathrm{DM_{obs}}(z)$} & {$\mathrm{DM_{MW,ISM}}(z)$} & Host & Ref. \\
\midrule
\endhead
\midrule
\multicolumn{6}{r}{\textit{Continued on next page}} \\
\endfoot
\bottomrule
\endlastfoot
\label{tab:frbs}
FRB 20200723B & 0.0085 & 244.0   & 33.0   & 3 & \cite{Shin_2024}        \\
FRB 20231229A & 0.019  & 198.5   & 65.13  & 3 & \cite{Amiri_2025}       \\
FRB 20240210A & 0.0237 & 283.73  & 28.7   & 3 & \cite{Shannon_2025}     \\
FRB 20231230A & 0.0298 & 131.4   & 32.23  & 3 & \cite{Amiri_2025}       \\
FRB 20181223C & 0.0302 & 111.61  & 19.9   & 3 & \cite{Bhardwaj_2024}    \\
FRB 20240201A & 0.0427 & 374.5   & 38.6   & 3 & \cite{Shannon_2025}     \\
FRB 20220207C & 0.0430 & 262.38  & 76.1   & 3 & \cite{Law_2024}         \\
FRB 20211127I & 0.0469 & 234.83  & 42.5   & 3 & \cite{Gordon_2023}      \\
FRB 20200223B & 0.0602 & 201.8   & 45.6   & 2 & \cite{Ibik_2024}        \\
FRB 20190303A & 0.064  & 223.2   & 29.8   & 2 & \cite{Michilli_2023}    \\
FRB 20231204A & 0.0644 & 221.0   & 34.94  & 2 & \cite{Amiri_2025}       \\
FRB 20231206A & 0.0659 & 457.7   & 37.51  & 3 & \cite{Amiri_2025}       \\
FRB 20231120A & 0.07   & 438.9   & 43.8   & 3 & \cite{Sharma_2024}      \\
FRB 20190418A & 0.0713 & 182.78  & 70.2   & 3 & \cite{Bhardwaj_2024}    \\
FRB 20211212A & 0.0715 & 206.0   & 27.1   & 3 & \cite{Gordon_2023}      \\
FRB 20231123A & 0.0729 & 302.1   & 37.11  & 3 & \cite{Amiri_2025}       \\
FRB 20231011A & 0.0783 & 186.3   & 58.55  & 3 & \cite{Amiri_2025}       \\
FRB 20220509G & 0.0894 & 269.53  & 55.6   & 3 & \cite{Law_2024}         \\
FRB 20230930A & 0.0925 & 456.0   & 70.0   & 3 & \cite{Anna_Thomas_2025} \\
FRB 20230124A & 0.0939 & 590.57  & 38.6   & 3 & \cite{Sharma_2024}      \\
FRB 20201124A & 0.098  & 413.52  & 139.9  & 2 & \cite{Ravi_2022}        \\
FRB 20230708A & 0.105  & 411.51  & 60.3   & 3 & \cite{Shannon_2025}     \\
FRB 20191106C & 0.1078 & 332.2   & 25.0   & 2 & \cite{Ibik_2024}        \\
FRB 20231128A & 0.1079 & 331.6   & 64.74  & 1 & \cite{Amiri_2025}       \\
FRB 20230222B & 0.11   & 187.8   & 57.0   & 3 & \cite{Amiri_2025}       \\
FRB 20220914A & 0.1139 & 631.28  & 54.7   & 3 & \cite{Law_2024}         \\
FRB 20190608B & 0.1178 & 338.7   & 37.3   & 3 & \cite{Chittidi_2021}    \\
FRB 20230703A & 0.1184 & 291.3   & 38.16  & 3 & \cite{Amiri_2025}       \\
FRB 20240213A & 0.1185 & 357.4   & 40.0   & 3 & \cite{Connor_2025}      \\
FRB 20230222A & 0.1223 & 706.1   & 33.31  & 3 & \cite{Amiri_2025}       \\
FRB 20190110C & 0.1224 & 221.6   & 37.1   & 2 & \cite{Ibik_2024}        \\
FRB 20230628A & 0.1265 & 344.95  & 39.0   & 3 & \cite{Sharma_2024}      \\
FRB 20240310A & 0.127  & 601.8   & 30.1   & 3 & \cite{Shannon_2025}     \\
FRB 20210807D & 0.1293 & 251.3   & 121.2  & 3 & \cite{Gordon_2023}      \\
FRB 20240114A & 0.13   & 527.65  & 49.7   & 1 & \cite{Tian_2024}        \\
FRB 20240209A & 0.1384 & 176.52  & 63.1   & 2 & \cite{Eftekhari_2025}   \\
FRB 20210410D & 0.1415 & 571.2   & 56.2   & 3 & \cite{Caleb_2023}       \\
FRB 20230203A & 0.1464 & 420.1   & 67.3   & 3 & \cite{Amiri_2025}       \\
FRB 20231226A & 0.1569 & 329.9   & 38.1   & 3 & \cite{Shannon_2025}     \\
FRB 20230526A & 0.157  & 361.4   & 31.9   & 3 & \cite{Shannon_2025}     \\
FRB 20220920A & 0.1582 & 314.99  & 39.9   & 3 & \cite{Law_2024}         \\
FRB 20200430A & 0.1608 & 380.1   & 27.2   & 3 & \cite{Heintz_2020}      \\
FRB 20210603A & 0.1772 & 500.15  & 39.5   & 3 & \cite{Cassanelli_2024}  \\
FRB 20220529A & 0.1839 & 246.0   & 40.0   & 1 & \cite{Li_2026}          \\
FRB 20230311A & 0.1918 & 364.3   & 67.24  & 3 & \cite{Amiri_2025}       \\
FRB 20220725A & 0.1926 & 290.4   & 30.7   & 3 & \cite{Shannon_2025}     \\
FRB 20121102A & 0.1927 & 557.0   & 188.4  & 1 & \cite{Chatterjee_2017}  \\
FRB 20221106A & 0.2044 & 343.8   & 34.8   & 3 & \cite{Shannon_2025}     \\
FRB 20240215A & 0.21   & 549.5   & 47.9   & 3 & \cite{Connor_2025}      \\
FRB 20230730A & 0.2115 & 312.5   & 114.65 & 3 & \cite{Amiri_2025}       \\
FRB 20210117A & 0.2145 & 729.1   & 34.4   & 3 & \cite{Bhandari_2023}    \\
FRB 20221027A & 0.229  & 452.5   & 47.2   & 3 & \cite{Sharma_2024}      \\
FRB 20191001A & 0.234  & 506.92  & 44.2   & 3 & \cite{Heintz_2020}      \\
FRB 20190714A & 0.2365 & 504.13  & 38.5   & 3 & \cite{Heintz_2020}      \\
FRB 20221101B & 0.2395 & 491.55  & 131.2  & 3 & \cite{Sharma_2024}      \\
FRB 20220825A & 0.2414 & 651.24  & 78.5   & 3 & \cite{Law_2024}         \\
FRB 20240304A & 0.2423 & 652.6   & 73.9   & 3 & \cite{Shannon_2025}     \\
FRB 20191228A & 0.2432 & 297.5   & 32.9   & 3 & \cite{Bhandari_2022}    \\
FRB 20231017A & 0.245  & 344.2   & 31.15  & 3 & \cite{Amiri_2025}       \\
FRB 20221113A & 0.2505 & 411.03  & 91.7   & 3 & \cite{Sharma_2024}      \\
FRB 20220307B & 0.2507 & 499.15  & 128.2  & 3 & \cite{Law_2024}         \\
FRB 20231123B & 0.2625 & 396.7   & 40.2   & 3 & \cite{Sharma_2024}      \\
FRB 20230307A & 0.271  & 608.9   & 37.6   & 3 & \cite{Sharma_2024}      \\
FRB 20221116A & 0.2764 & 643.45  & 132.3  & 3 & \cite{Sharma_2024}      \\
FRB 20220105A & 0.2785 & 583.0   & 22.0   & 3 & \cite{Gordon_2023}      \\
FRB 20210320C & 0.2797 & 384.8   & 39.3   & 3 & \cite{Shannon_2025}     \\
FRB 20221012A & 0.2847 & 441.08  & 54.3   & 3 & \cite{Law_2024}         \\
FRB 20240229A & 0.287  & 491.15  & 38.0   & 3 & \cite{Connor_2025}      \\
FRB 20190102C & 0.2912 & 364.5   & 57.4   & 3 & \cite{Bhandari_2020}    \\
FRB 20220506D & 0.3004 & 396.97  & 84.6   & 3 & \cite{Law_2024}         \\
FRB 20230501A & 0.301  & 532.5   & 125.6  & 3 & \cite{Sharma_2024}      \\
FRB 20180924B & 0.3214 & 361.42  & 40.5   & 3 & \cite{Bannister_2019}   \\
FRB 20231025B & 0.3238 & 368.7   & 65.58  & 3 & \cite{Amiri_2025}       \\
FRB 20230626A & 0.327  & 451.2   & 39.2   & 3 & \cite{Sharma_2024}      \\
FRB 20180301A & 0.3304 & 552.0   & 151.7  & 1 & \cite{Bhandari_2022}    \\
FRB 20231220A & 0.3355 & 491.2   & 49.9   & 3 & \cite{Connor_2025}      \\
FRB 20211203C & 0.3439 & 636.2   & 63.7   & 3 & \cite{Gordon_2023}      \\
FRB 20230808F & 0.3472 & 653.2   & 31.0   & 3 & \cite{Hanmer_2025}      \\
FRB 20220208A & 0.351  & 437.0   & 101.6  & 3 & \cite{Sharma_2024}      \\
FRB 20220726A & 0.361  & 686.55  & 89.5   & 3 & \cite{Sharma_2024}      \\
FRB 20230902A & 0.3619 & 440.1   & 34.1   & 3 & \cite{Shannon_2025}     \\
FRB 20220717A & 0.363  & 637.34  & 118.3  & 3 & \cite{Rajwade_2024}     \\
FRB 20200906A & 0.3688 & 577.8   & 35.8   & 3 & \cite{Bhandari_2022}    \\
FRB 20220330D & 0.3714 & 468.1   & 38.6   & 3 & \cite{Sharma_2024}      \\
FRB 20240119A & 0.376  & 483.1   & 38.0   & 3 & \cite{Connor_2025}      \\
FRB 20190611B & 0.3778 & 321.4   & 57.8   & 3 & \cite{Heintz_2020}      \\
FRB 20220501C & 0.381  & 449.5   & 30.6   & 3 & \cite{Shannon_2025}     \\
FRB 20230506C & 0.3896 & 772.0   & 68.0   & 2 & \cite{Anna_Thomas_2025} \\
FRB 20220204A & 0.4012 & 612.58  & 50.7   & 3 & \cite{Sharma_2024}      \\
FRB 20230712A & 0.4525 & 587.57  & 39.2   & 3 & \cite{Sharma_2024}      \\
FRB 20181112A & 0.4755 & 589.27  & 41.7   & 3 & \cite{Prochaska_2019}   \\
FRB 20220310F & 0.478  & 462.24  & 46.3   & 3 & \cite{Law_2024}         \\
FRB 20220918A & 0.491  & 656.8   & 153.1  & 3 & \cite{Shannon_2025}     \\
FRB 20190711A & 0.522  & 593.1   & 56.5   & 1 & \cite{Heintz_2020}      \\
FRB 20230216A & 0.531  & 828.0   & 38.5   & 3 & \cite{Sharma_2024}      \\
FRB 20221219A & 0.553  & 706.71  & 44.4   & 3 & \cite{Sharma_2024}      \\
FRB 20230814A & 0.553  & 696.4   & 104.8  & 3 & \cite{Connor_2025}      \\
FRB 20190614D & 0.6    & 959.2   & 87.8   & 3 & \cite{Law_2024}         \\
FRB 20220418A & 0.622  & 623.25  & 36.7   & 3 & \cite{Law_2024}         \\
FRB 20190523A & 0.66   & 760.8   & 37.2   & 3 & \cite{Ravi_2019}        \\
FRB 20240123A & 0.968  & 1462.0  & 90.2   & 3 & \cite{Connor_2025}      \\
FRB 20221029A & 0.975  & 1391.75 & 43.8   & 3 & \cite{Sharma_2024}      \\
FRB 20220610A & 1.016  & 1458.15 & 31.0   & 3 & \cite{Ryder_2023}       \\
FRB 20220521B & 1.354  & 1342.9  & 138.8  & 3 & \cite{Shannon_2025}     \\
\end{longtable}


\section{Interpolation Grid Validation}
\label{app:grid}

For each $f(R)$ model, we validate the four-dimensional interpolation grid 
by comparing $H(z)$ obtained from direct ODE integration with the interpolated 
values at 300 random interior points drawn uniformly in the parameter space
$(H_0,\Omega_{\rm m},\theta_{f(R)})$, excluding the outermost grid nodes to
avoid boundary artifacts.

At each test point, we evaluate the relative interpolation error,
\begin{equation}
\Delta_H(z)=
\left|
\frac{H_{\rm interp}(z)-H_{\rm ODE}(z)}
{H_{\rm ODE}(z)}
\right|,
\end{equation}
and record its maximum value over the redshift interval $z\in[0,4]$.
The resulting distribution of maximum errors is summarized in
Table~\ref{tab:grid_validation} through the median, the 95th percentile, 
and the global maximum.

\begin{table}[h!]
\centering
\begin{tabular}{c c c c}
\toprule
\textbf{Model} & \textbf{Median [\%]} & \textbf{95th pct [\%]} & \textbf{Max error [\%]} \\
\midrule
AB  & 0.0062 & 0.0307 & 0.0492 \\
HS  & 0.0053 & 0.0293 & 0.1029 \\
ST  & 0.0055 & 0.0274 & 0.0503 \\
\bottomrule
\end{tabular}
\caption{Summary of interpolation errors for $H(z)$ in each $f(R)$ model. 
Errors correspond to the distribution of the maximum relative deviation 
over $z \in [0,4]$ evaluated at 300 random interior points.}
\label{tab:grid_validation}
\end{table}

All models satisfy the accuracy requirement adopted in Sec.~\ref{sec:numerics},
with maximum relative interpolation errors below $0.5\%$ throughout the parameter
space. When propagated to the observables at the data redshifts, the resulting 
deviations remain negligible compared to the observational uncertainties.
The corresponding impact on the likelihood is entirely subdominant, with upper
bounds on the induced shifts satisfying $\Delta\chi^2 \lesssim 3 \times 10^{-2}$
for all models and datasets. We therefore conclude that the interpolation procedure
is numerically robust and does not affect the statistical inference or the model 
comparison results. Additional implementation details, validation tests, and 
numerical benchmarks are available in the public repository cited previously.


\bibliographystyle{JHEP}
\bibliography{biblio.bib}

\end{document}